\documentclass[a4paper,structabstract]{aa}

\usepackage{aas_macros}
\usepackage{natbib}
\usepackage{txfonts}
\usepackage{array}
\usepackage{wasysym} 
\usepackage{mathtools} 
\usepackage{nicefrac}
\usepackage[rightcaption]{sidecap}
\sidecaptionvpos{figure}{m}

\usepackage{mathptmx}
\usepackage{hyperref}

\hypersetup{
   colorlinks=true,
   citecolor=blue,
   linkcolor=blue,
   urlcolor=black
   }

\usepackage{graphicx,subfigure}

\def\aap{{\it Astron. \& Astrophys}}

\usepackage{color}

\begin{document}

\title{From thermal dissociation to condensation in the atmospheres of ultra hot Jupiters: WASP-121b in context}
\titlerunning{Thermal dissociation and condensation in the atmospheres of ultra hot Jupiters}
\authorrunning{Parmentier et al.}

\author{Vivien Parmentier\inst{1}
\and
Mike R. Line\inst{2}
\and
Jacob L. Bean\inst{3}
\and
Megan Mansfield\inst{4}
\and
Laura Kreidberg\inst{5,6}
\and
Roxana Lupu\inst{7}
\and 
Channon Visscher\inst{8,9}  
\and
Jean-Michel D\'esert\inst{10} 
\and
Jonathan J. Fortney\inst{11} 
\and 
Magalie Deleuil\inst{1}
\and
Jacob Arcangeli\inst{10}
\and
Adam P. Showman\inst{12}
\and
Mark S. Marley\inst{13}
}

\offprints{V. Parmentier}

\institute{Aix Marseille Univ, CNRS, LAM, Laboratoire d'Astrophysique de Marseille, Marseille, France\\
           \email{vivien.parmentier@lam.fr}
           \and 
School of Earth \& Space Exploration, Arizona State University, Tempe AZ 85287, USA
\and 
Department of Astronomy \& Astrophysics, University of Chicago, 5640 S. Ellis Avenue, Chicago, IL 60637, USA
\and
Department of Geophysical Sciences, University of Chicago, 5734 S. Ellis Avenue, Chicago, IL 60637, USA
\and
Harvard-Smithsonian Center for Astrophysics, 60 Garden Street, Cambridge, MA 02138
\and
Harvard Society of Fellows, 78 Mount Auburn Street, Cambridge, MA 02138
\and
BAER Institute, NASA Research Park, Moffett Field, CA 94035
\and
Chemistry \& Planetary Sciences, Dordt College, Sioux Center IA, USA
\and
Space Science Institute, Boulder CO, USA
\and
Anton Pannekoek Institute for Astronomy, University of Amsterdam, Science Park 904, 1098 XH Amsterdam, The Netherlands
\and
Department of Astronomy and Astrophysics, University of California, Santa Cruz, CA 95064
\and
Department of Planetary Sciences, Lunar and Planetary Laboratory, University of Arizona, Tucson AZ, USA
\and
NASA Ames Research Center
Moffett Field, CA 94035, USA
}


\abstract
{A new class of exoplanets has emerged: the ultra hot Jupiters, the hottest close-in gas giants. The majority of them have weaker-than-expected spectral features in the $1.1-1.7\mu m$ bandpass probed by HST/WFC3 but stronger spectral features at longer wavelengths probed by Spitzer. This led previous authors to puzzling conclusions about the thermal structures and chemical abundances of these planets.
}
{We investigate how thermal dissociation, ionization, $\rm H^-$ opacity, and clouds shape the thermal structures and spectral properties of ultra hot Jupiters.
}
{We use the SPARC/MITgcm to model the atmospheres of four ultra hot Jupiters and discuss more thoroughly the case of WASP-121b. We expand our findings to the whole population of ultra hot Jupiters through analytical quantification of the thermal dissociation and its influence on the strength of spectral features.
}
{ We predict that most molecules are thermally dissociated and alkalies are ionized in the dayside photospheres of ultra hot Jupiters. This includes $\rm H_2O$, TiO, VO, and $\rm H_2$ but not CO, which has a stronger molecular bond. The vertical molecular gradient created by the dissociation significantly weakens the spectral features from $\rm H_2O$ while the $4.5 \mu m$ CO feature remains unchanged. The water band in the HST/WFC3 bandpass is further weakened by the continuous opacity of the $\rm H^-$ ions. Molecules are expected to recombine before reaching the limb, leading to order of magnitude variations of the chemical composition and cloud coverage between the limb and the dayside.
 }
{Molecular dissociation provides a qualitative understanding of the lack of strong spectral features of water in the $1-2\mu m$ bandpass observed in most ultra hot Jupiters. Quantitatively, our model does not provide a satisfactory match to the WASP-121b emission spectrum. Together with WASP-33b and Kepler-33Ab, they seem the outliers among the population of ultra hot Jupiters, in need of a more thorough understanding.}

\maketitle
%
%
\section{Introduction}

Ultra hot Jupiters, which we define as gas giants with dayside temperatures $\gtrsim$\,2,200\,K, have recently emerged as a population of exoplanets with distinct atmospheric characteristics \citep{Arcangeli2018,Bell2018}. The hottest of the hot Jupiters are the best targets for thermal emission measurements because they are the easiest to observe due to their favorable contrast with their host stars. They also provide an opportunity to study the physics and chemistry of planetary atmospheres in a regime that is far removed from the conditions of the planets in the solar system. Ultra hot Jupiters are actually a rare outcome of planet formation \citep{wright2012}, yet fortunately many are known mainly due to the excellent sky coverage of ground-based transit surveys like WASP \citep{Pollacco2006}, HATNet and HATSouth \citep{Bakos2018}, KELT \citep{Pepper2007}, and MASCARA \citep{Snellen2013}.

Thermal emission measurements of ultra hot Jupiters using the WFC3 instrument on the Hubble Space
Telescope (HST) have been primarily motivated by the chance to measure water abundances to understand the bulk composition of the planets and the hope that it can be used to trace their formation mechanisms \citep[e.g.,][]{Kreidberg2014a}. However, the shape of the emission spectrum in the WFC3 bandpass ($\approx 1.1-1.7\mu m$) depends both on the water abundance and the thermal structure of the planet. In an atmosphere that is chemically homogeneous, an isothermal structure will exhibit a blackbody spectrum, a thermal structure with increasing temperature with increasing pressure ("non-inverted temperature profile") will exhibit a spectrum with absorption features, and a thermal structure with decreasing temperature with increasing pressure ("inverted temperature profile") will exhibit a spectrum with emission features~\citep[e.g.,][]{Fortney2008,Madhusudhan2010,Line2016a}. Therefore the atmospheric thermal structure of a planet has a major influence on the ease with which chemical compositions can be measured from emission spectroscopy.

The currently available set of HST/WFC3 thermal emission measurements for hot Jupiters display a variety of spectral morphology. The presence of a non-inverted thermal structure has been detected with high certainty in the two relatively cool hot Jupiters WASP-43b \citep{Stevenson2014b} and HD\,209458b \citep{Line2016a}, which is consistent with theoretical expectations \citep{Fortney2008}. For hotter planets, however, there is a surprising prevalence of blackbody-like emission spectra with no clear absorption or emission features (e.g., WASP-12b: \citet{Swain2013}; WASP-103b: \citet{Cartier2017,Kreidberg2018}; WASP-18b: \citet{Arcangeli2018}; and HAT-P-7b: \cite{Mansfield2018}). Despite the prevalence of blackbody-like spectra among the hottest planets, the spectra for some of these objects have been interpreted as showing emission features (e.g., WASP-33b: \citet{Haynes2015} and WASP-121b: \citet{Evans2017}), while in one case an absorption feature was reported \citep[Kepler-13Ab,][]{Beatty2017a}. 

Obtaining a self-consistent solution explaining the features seen in some of these hot planet atmospheres and the lack of spectral features seen in others has been a challenge to theorists. Most of the latest studies have been based on retrieval modeling and point toward a large diversity in both the chemical composition and the thermal structure of the hottest hot Jupiters. Isothermal \citep[e.g., WASP-103b,][]{Delrez2018}), inverted \citep[e.g., WASP-121b,][]{Evans2017}, and non-inverted pressure-temperature (PT) profiles \citep[e.g.,][]{Beatty2017a} together with compositions ranging from a sub-solar metallicity and oxygen-rich composition \citep[e.g., WASP-33b,][]{Haynes2015} to a super-solar metallicity and carbon-enriched composition \citep[e.g., WASP-18b,][]{Sheppard2017} have been claimed based on emission spectra obtained with both HST/WFC3 and Spitzer/IRAC. Although useful to quantify the atmospheric abundances when all the relevant physics is incorporated, atmospheric retrievals can, however, be biased toward incorrect solutions if some important physical processes are missing~\citep{Line2016,Feng2016}. 

Here we discuss the important roles of two physical effects that have been neglected in most previous atmospheric retrieval studies trying to constrain molecular abundances from the emission spectra of ultra hot Jupiters: $\rm H^-$ opacities and the thermal dissociation of water. No new physics is discovered here, since both effects have been taken into account in the study of stellar atmospheres for several decades~\citep{Pannekoek1931,Wildt1939,Chandrasekhar1945}. Moreover, since most self-consistent one-dimensional (1D) radiative/convective models used in the study of exoplanet atmospheres are legacy models from stellar and substellar studies, they almost always have these two effects already incorporated~\citep{Marley1996,Burrows1997,Barman2001,Hubeny2003,Fortney2008,Lothringer2018}. More recent forward models that have been built specifically with modeling exoplanet spectra in mind have usually taken into account molecular dissociation but have often neglected the presence of $\rm H^-$ opacities~\citep{Tremblin2015,Baudino2015}, except in a few cases~\citep[e.g.,][]{Molliere2015}. Retrieval models, used to quantify molecular abundances from spectra often neglect both $\rm H^-$ opacities and the height dependence of the molecular abundances~\citep[e.g.,][]{Madhusudhan2009, Lee2012, Line2013}. More recently, retrieval models have begun incorporating more thorough forward models including chemical equilibrium calculations~\citep[e.g.,][]{Benneke2015,Kreidberg2015} and $\rm H^-$ opacities~\citep{Arcangeli2018}. 

The present paper provides a framework for understanding the physical processes that go into shaping the emergent spectra of ultra hot Jupiters. We first describe our modeling approach in Sect.~\ref{sec:Models}. Then, in Sect.~\ref{sec:Emissions}, we use a combination of global circulation models and analytical models to study the effects that water dissociation and $\rm H^-$ opacities can have on the emission spectrum of WASP-121b and other ultra hot Jupiters. In Sect.~\ref{Sec:3DStruct} we discuss the thermo-chemical three-dimensional (3D) structure of these planets and re-interpret the transmission spectrum of WASP-121b by showing that some molecules can recombine and even condense at the atmospheric limb. Finally, we expand our findings to the whole population of planets and discuss the link with M dwarf analogs in Sect.~\ref{sec:pop}. This study is complementary to other articles \citep{Arcangeli2018, Mansfield2018, Kreidberg2018} that more thoroughly interpret the emission spectra of individual planets within the context of a retrieval framework that includes the aforementioned physics.

\section{Model}
\label{sec:Models}
\subsection{Global circulation model}
We use the SPARC/MITgcm global circulation model~\citep{Showman2009} to model the atmosphere of WASP-121b. The model solves the primitive equations on a cubic-sphere grid. It has been successfully applied to a wide range of hot Jupiters~\citep{Showman2009,Kataria2015,Kataria2016,Lewis2017,Parmentier2013,Parmentier2016}, including a few ultra hot Jupiters~\citep{Zhang2018,Kreidberg2018} but no in-depth analysis of the simulations has yet been performed for planets hotter than WASP-19b~\citep[\rm T$_{\rm eq}$=2050K,][]{Kataria2016}.

Our pressure ranges from 200 bar to 2 $\mu$bar over 53 levels, so that we have a resolution of almost three levels per scale height. We use a horizontal resolution of C32, equivalent to an approximate resolution of 128 cells in longitude and 64 in latitude and a time-step of 25 s. We initialize the model at rest with a temperature profile from the analytical model of~\citet{Parmentier2015} that uses the analytical expression of~\citet{Parmentier2014} adjusted to represent the global average temperature profile of solar-composition atmospheres~\citet{Fortney2007}. The simulations all run for 300 Earth days, a length of time over which the observable atmosphere has reached a quasi steady-state~\citep{Showman2009}. The first 200 days of the simulation are then discarded and the pressure-temperature profiles are averaged over the last 100 days of simulation.

We assume that the specific heat capacity, heat capacity ratio, and the mean molecular weight of the atmosphere are constant and equal to, respectively, $C_{\rm p}=1.3\times10^4\, J\, kg^{-1}\, K^{-1}$, $\gamma=1+2/7$ and $\mu=2.3m_{\rm H}$, appropriate for a $\rm H_2$-dominated atmosphere. This assumption, however, reaches its limit for the hot planets considered here where $\rm H_2$ can dissociate. In the regions dominated by $\rm H$ instead of $\rm H_2$, $C_p$ should be twice as big, $\mu$ twice as small, and $\gamma$ should become equal to $1+2/3$~\citep{Showman2002}. This variation is taken into account in the radiative transfer calculations but not in the dynamical equations. Although the variation in the mean molecular weight and heat capacity of the atmosphere is likely to play an important role in determining the atmospheric dynamics of the atmosphere~\citep{Zhang2017}, it is usually a second-order one compared to the radiative effect associated with the changes in the dominant molecule~\citet{Drummond2018}.

\subsection{Radiative transfer model}
Radiative transfer is handled both in the 3D simulations and during the post-processing using the plane-parallel radiative transfer code of \cite{Marley1999}. The code was first developed for Titan's atmosphere \citep{McKay1989} and since then has been extensively used for the study of giant planets \citep{Marley1996}, brown dwarfs \citep{Marley2002,Burrows1997}, and hot Jupiters \citep{Fortney2005, Fortney2008, Showman2009}. We use the opacities described in \cite{Freedman2008}, including more recent updates \citep{Freedman2014}, and the molecular abundances described in Sect.~\ref{sec:Abundances}. Our opacity database do not include atomic species such as Fe and Mg although it has recently been shown that they can contribute significantly to the opacities in these ultra hot atmospheres~\citep[e.g.][]{Lothringer2018}.

The version of the code we employ solves the radiative transfer equation in the two-stream approximation using the delta-discrete ordinates method of \cite{Toon1989} for the incident stellar radiation and the two-stream source function method, also of \cite{Toon1989}, for the thermal radiative transfer. Molecular and atomic opacities are treated using  the correlated-k method \citep{Goody1989}: the spectral dimension is divided into a number of bins and within each bin the information of typically 10,000 to 100,000 frequency points is compressed inside a single cumulative distribution function that is then interpolated using eight $k$-coefficients. For the gas, Rayleigh scattering is taken into account in the calculation together with collision induced absorption and $\rm H^-$ continuum opacities. { The radiative effect of clouds is neglected in the simulations unless otherwise stated.}%

Our radiative transfer calculations can be done with two different spectral resolutions. When coupled to the GCM, we use 11 frequency bins that have been carefully chosen to maximize the accuracy and the speed of the calculation~\citep{Kataria2013}. We then use the thermal structure outputted from the global circulation model and post-process it with a higher-resolution version of our radiative transfer model (196 frequency bins ranging from $0.26$ to $300\mu m$). For this, we solve the two-stream radiative transfer equations along the line of sight for each atmospheric column and for each planetary phase considering absorption, emission, and scattering.  This method, similar to the calculation of~\citet{Fortney2006a} and~\citet{Parmentier2016}, naturally takes into account geometrical effects such as limb darkening. The stellar flux is assumed to be a collimated flux propagating in each atmospheric column with an angle equal to the angle between the local vertical and the direction of the star.

Although in general the line-by-line opacities used to calculate the k-coefficient tables in the post-processing are the same as the ones used in the global circulation,  to test the influence of a given species on the spectrum, we can run the radiative transfer post-processing assuming a different chemical composition for the atmosphere. 

{ To calculate the transmission spectrum, we use a simpler procedure than~\citet{Fortney2010}. We first calculate the height difference between the $1\,\rm bar$ radius, assumed constant with longitude and latitude, and the transit radius at a given wavelength. For a given azimuth\footnote{We define the azimuth as the projected angle between the substellar point, the east limb, and the point of interest; see also Fig.~\ref{fig:Transmission}.} at the limb, a given layer, and a given k-coefficient bin, we calculate the chord optical depth by multiplying the vertical optical depth of the layer (a natural output of our radiative transfer model) by the geometric factor $\sqrt(2\pi Rp/H)$~\citep[see Eq. 6 of][]{Fortney2005b}. We then perform the weighed sum of the horizontal optical depth for each k-coefficient to obtain the horizontal optical depth in each wavelength bin. We repeat this operation for every layer to find the radius of the layer where this part of the limb becomes opaque (corresponding to a horizontal optical depth $\tau\approx0.56$ for a hot Jupiter; see  ~\citet{LecavelierDesEtangs2008} and~\citet{deWit2013}). We repeat this operation for every slice of the limb, considering the change in scale height, chemical composition, and cloud abundance due to the local temperature-pressure profile. We then average the radius calculated at each different limb azimuth to obtain the mean transit radius.} Finally we adjust the 1 bar radius within the measured uncertainties to obtain the best match of the data. { This method assumes that the light is mostly absorbed near the layer where the atmosphere becomes opaque and ignores the fact that light traveling through the limb crosses layers at different longitude that might have different properties}. Despite this approximation, our method gives similar results for hot Jupiters to the more complex method used by~\citet{Fortney2010}.

{ In one instance, we post-process our cloudless model by assuming the presence of condensates. In a given model cell, we assume that all condensable material is trapped into condensate of size $a$, with $a$ being a free parameter. The number of particles is calculated by assuming mass conservation in a given cell and that neither vertical mixing nor settling is at play. This approach effectively corresponds to the case of very large vertical mixing strength. The absorption opacity, the single scattering albedo, and the asymmetry parameter are determined with the Mie theory. More details can be found in~\citet{Parmentier2016}.}

\begin{figure}[h!]
\includegraphics[width=\linewidth]{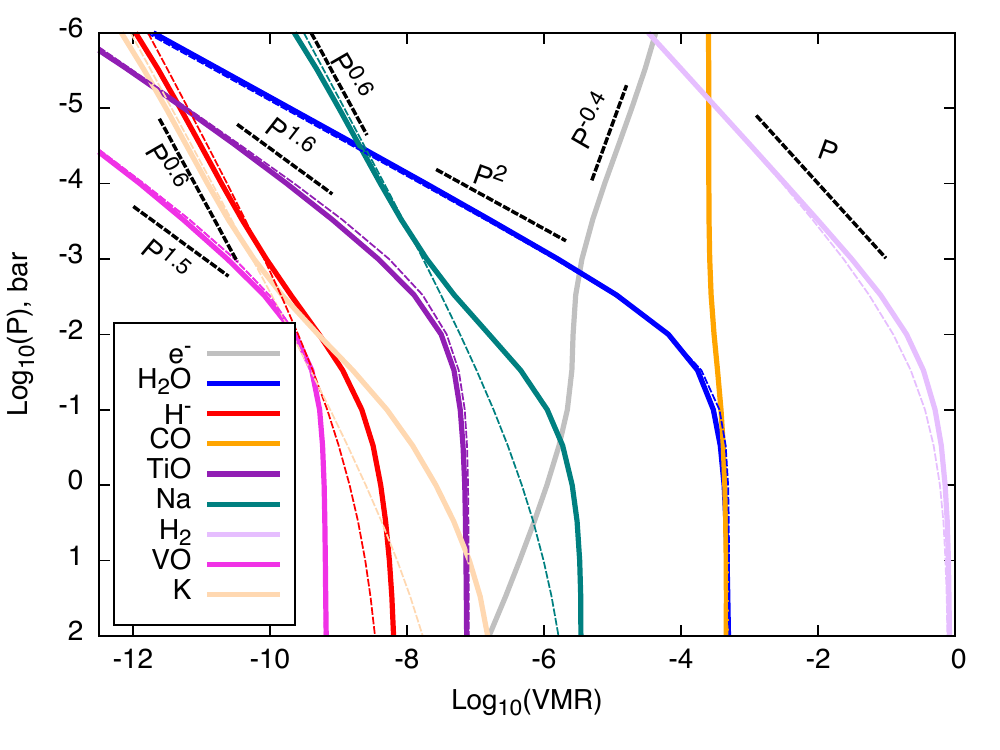}
\caption{Abundances of the main spectrally active species in hot atmospheres as a function of pressure and for a temperature of $3000\,\rm K$. The solid lines are calculated with the NASA CEA Gibbs minimization code (see text for details) while the dashed lines are the analytical fits from Eq.~\ref{eq:Abunds} using the coefficients of Table~\ref{Table:alpha}.}
\label{fig:Abundances}
\end{figure}
\subsection{Molecular abundances}
\label{sec:Abundances}
Molecular and atomic abundances are calculated using a modified version of the NASA CEA Gibbs minimization code \citep[see][]{gordon1994} as part of a model grid previously used to explore gas and condensate equilibrium chemistry in substellar objects \citep[][]{Moses2013,Skemer2016,Kataria2015,Wakeford2017,Burningham2017,Marley2017DPS} over a wide range of atmospheric conditions.  Here we assume solar elemental abundances { and local chemical equilibrium with rainout of condensate material, meaning that the gas phase does not interact with the solid phase}. As in previous thermochemical grid models \citep[see][]{Lodders2002a,Visscher2006,Visscher2010}, molecular species thermally dissociate at high temperatures and low pressures via net thermochemical reactions such as
\begin{align*}
\textrm{H}_2 & \rightleftarrows 2\textrm{H}\\
\textrm{H}_2\textrm{O} & \rightleftarrows 2\textrm{H} + \textrm{O}\\
\textrm{Ti}\textrm{O} & \rightleftarrows \textrm{Ti} + \textrm{O}\\
\textrm{V}\textrm{O} & \rightleftarrows \textrm{V} + \textrm{O},
\end{align*}
and the abundances of molecules such as $\rm H_2$, H$_{2}$O, TiO and VO decrease as they dissociate into their atomic components (see Fig.~\ref{fig:Abundances}). The  model also considers thermal ionization of key atmospheric species (favored at high temperatures and low pressures) via net thermochemical reactions including
\begin{align*}
\textrm{Na} & \rightleftarrows \textrm{Na}^+ + \textrm{e}^-\\
\textrm{K} & \rightleftarrows \textrm{K}^+ + \textrm{e}^-\\
\textrm{Ti} & \rightleftarrows \textrm{Ti}^+ + \textrm{e}^-\\
\textrm{V} & \rightleftarrows \textrm{V}^+ + \textrm{e}^-, 
\end{align*}
lowering the abundances of neutral atomic species such as Na, K,  Ti and V  (see Fig.~\ref{fig:Abundances} and also \citet{Lodders2002}) with decreasing pressure.  In addition, we consider ionization reactions such as
\begin{align*}
\textrm{H} + \textrm{e}^- & \rightleftarrows \textrm{H}^-,
\end{align*}
producing species such as H$^-$ that also are pressure-dependent (although in this case, the abundance of the ion H$^-$ decreases with lower pressures).  

As shown in Fig~\ref{fig:Abundances}, for $P<100\,\rm mbar$, the pressure dependence of the abundances can be represented by a power law in the form $VMR\propto P^\alpha$ with the values of $\alpha$ given in Table~\ref{Table:alpha}. Carbon monoxide is a notable exception: due to its much stronger triple bond, the dissociation of CO (via CO $\rightleftarrows$ C + O) requires higher temperatures \citep[see][]{Lodders2002a}. More generally, we find that the volume mixing ratios of several elements can be approximated by the relationship:
\begin{equation}
\frac{1}{A}=\left(\frac{1}{A_{0}^{0.5}}+\frac{1}{A_{d}^{0.5}}\right)^2,\label{eq:Abunds}
\end{equation}
where $A_{0}$ is the deep abundance, unaffected by dissociation and $A_{d}$ is the abundance in the region dominated by dissociation. $A_{d}$ can be well approximated by :
\begin{equation}
\log_{10}{A_{d}} \approx \alpha \log_{10}{P}+\frac{\beta}{T}+\gamma 
\label{eq:AbundsD}
,\end{equation}
where the coefficients $\alpha$, $\beta$ and $\gamma$ are given in Table~\ref{Table:alpha}.  As the abundances of different species follow a different power law, the abundance ratio between these species varies throughout the atmosphere. In particular, we expect the \emph{relative} abundances of optical absorbers such as TiO, VO and Na to increase with decreasing pressure compared to the abundance of H$_{2}$O, the main component responsible for radiative cooling (e.g., see Fig~\ref{fig:Abundances}). From planet to planet, the depth of the photosphere should change depending on planet-specific parameters such as gravity and metallicity, leading to a change in the molecular abundance ratio observed at the photosphere.

\begin{table}
\centering
\caption{Coefficients for approximating atomic and molecular abundances following Eqs.~\ref{eq:Abunds} and~\ref{eq:AbundsD}$^{1}$}
\label{Table:alpha}
\begin{tabular}{|c | c | c | c | c | c }
\hline 
Species & $\alpha$ & $10^{-4}\beta$ & $\gamma$ & $\log_{10}{A_0}$\\
\hline
$\rm H_2$ & 1 & 2.41 & 6.5 & -0.1\\
$\rm H_2O$ & 2 & 4.83 & 15.9 & -3.3  \\
$\rm TiO$ & 1.6 & 5.94 & 23.0 & -7.1  \\
$\rm VO$ & 1.5 & 5.40 & 23.8 & -9.2  \\
$\rm H^-$ & 0.6 &-0.14 & 7.7 & -8.3 \\
$\rm Na $ & 0.6 & 1.89 & 12.2 & -5.5 \\
$\rm K $ & 0.6 & 1.28 & 12.7 & -7.1 \\
\hline
\end{tabular}
\\
$^{1}$$1/\sqrt{A}=1/\sqrt{A_0}+1/\sqrt{A_{d}}$ with $A_{d}=10^{-\gamma} P^\alpha10^{\beta/T}$\\
P in bar, T in Kelvin.\\
Valid for a solar-composition gas with $2000\,\rm K<T<4000\,\rm K$ and $200\,\rm bar<P<1\mu\,\rm bar$.
\end{table}

\section{How molecular dissociation and $\rm H^-$ opacities shape the emission spectrum of WASP-121b}
\label{sec:Emissions}
\subsection{Emission spectrum}
\label{Sec:EmSpec}
\begin{figure}[h!]
\includegraphics[width=\linewidth]{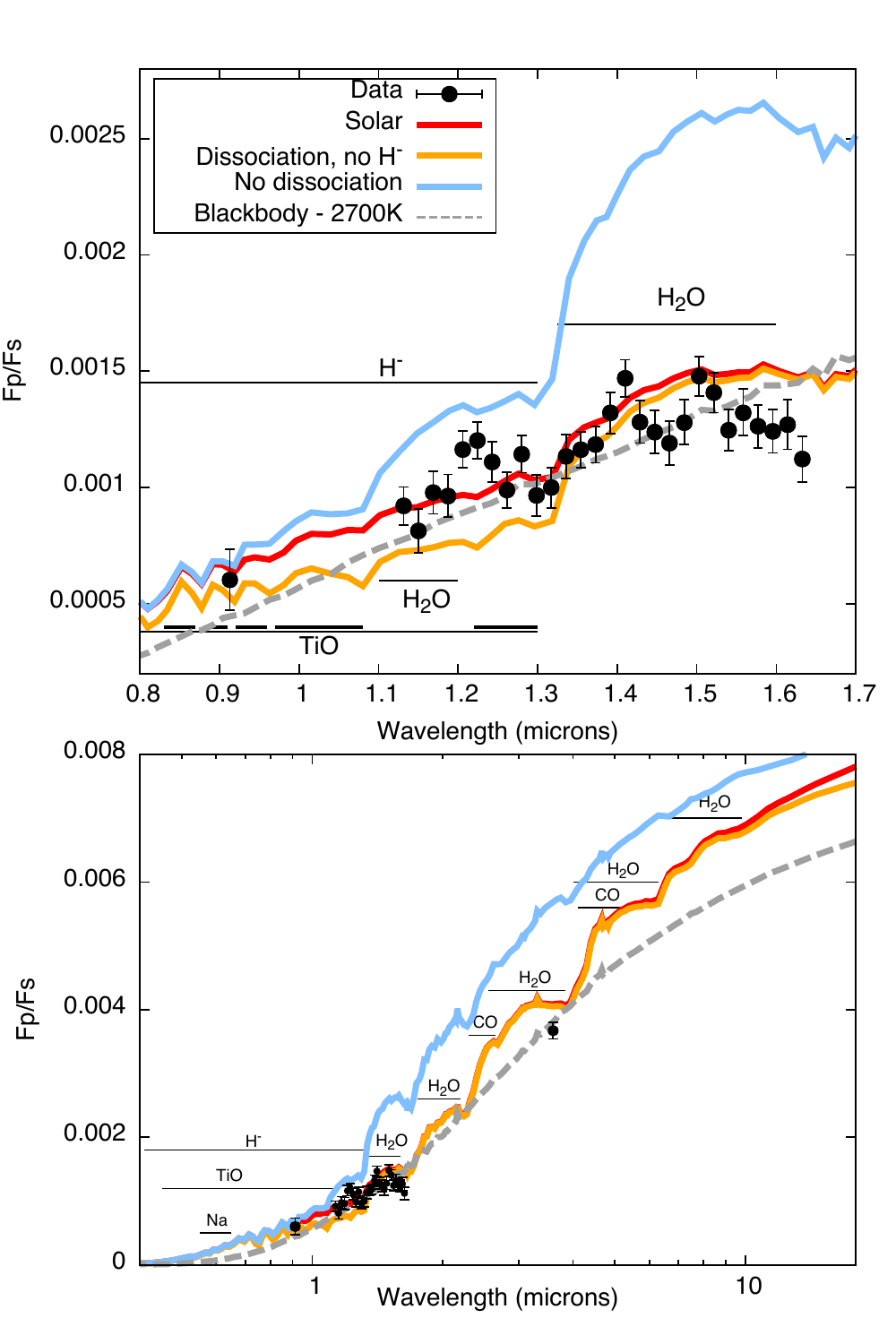}
\caption{Emission spectrum of WASP-121b obtained from Trappist (at $0.9\mu m$, see~\citet{Delrez2016}), HST/WFC3 (between 1 and $2\mu m$) and Spitzer (at $3.6\mu m$, see~\citet{Evans2017}). The observations are compared to the dayside spectrum from a SPARC/MITgcm with solar metallicity. Also shown are models with the same thermal structure but with either $\rm H^-$ opacities or water dissociation neglected when calculating the spectrum. Both $\rm H^-$ opacities and water dissociation are responsible for the weak water signature in the WFC3 bandpass.}
\label{fig:Spectrum}
\end{figure}

\begin{figure*}
\sidecaption
\includegraphics[width=10cm]{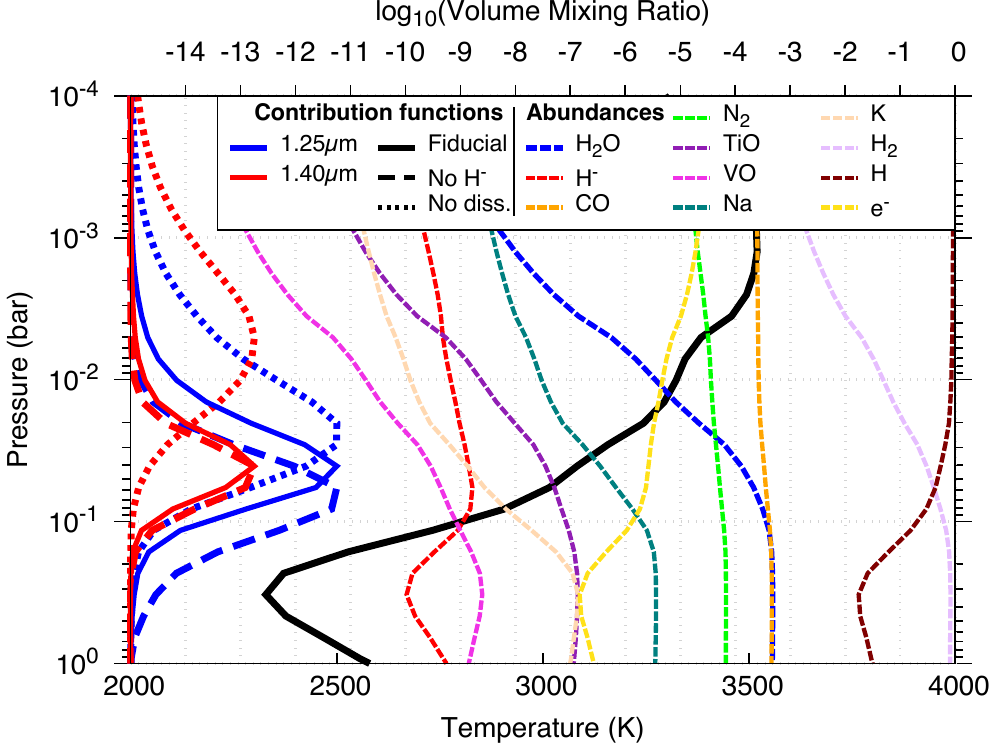}
\caption{Structure of WASP-121b at the substellar point. We plot the pressure-temperature profile (black line) together with elemental abundances (dashed lines). Contribution functions at $1.25\mu m$ (blue) and $1.4\mu m$ (red) are plotted on the left side of the plot for the fiducial model (solid line), the model post processed without $\rm H^-$ opacities (long dashed line) and the model post processed with a constant water abundance (dotted line). Contribution functions have been normalized to go from zero to one and scaled to improve visibility.}
\label{fig:ContrFunctions}
\end{figure*}

WASP-121b is a $M=1.18 M_{\rm J}$, $R=1.86R_{\rm J}$ planet orbiting at a distance of $0.0254\,\rm AU$ from an F6 star ($R_{\rm s}=1.45R_{\rm Sun}$,$T_{\rm eff}=6460\,\rm K$). Given the high temperature of the planet, the brightness of its host star (V=10.4) and the low planet gravity ($g=8.5m/s^2$), it is a prime candidate for atmospheric characterization. The transit of the planet was observed from ground-based observatories~\citep[Trappist and EulerCam, see][]{Delrez2016} and with HST/WFC3~\citep{Evans2016}, whereas the eclipse of the planet was observed with HST/WFC3, Trappist, and Spitzer~\citep{Evans2017}.    

WASP-121b is part of a group of ultra hot Jupiters that exhibit smaller spectral features in the $1.1-1.7\mu m$ wavelength range than their cooler analogs (see Sect.~\ref{sec:Obs} for a comparison). As shown in Fig.~\ref{fig:Spectrum}, the spectrum of WASP-121b lacks strong emission features and, although the observations rule out a blackbody emission spectrum~\citep{Evans2017}, only a few datapoints are more than two sigma away from the best fit blackbody model. The lack of a strong water feature in the emission spectrum of WASP-121b has been puzzling. The four reddest points of the WFC3 spectrum, in particular, point toward a non-isothermal atmosphere. If interpreted as being part of a water emission feature, one would expect a large flux difference between the $1.2-1.35\mu m$ region (outside the water band) and the $1.35-1.5\mu m$ region (inside the water band). The lack of such a feature led~\citet{Evans2017} to conclude the presence of an unknown absorber in the $1.2-1.35\mu m$ region that would fill the gap between the water bands. Vanadium oxide was proposed to be this absorber, { but the lack of TiO emission bands (while TiO is supposed to be 100x more abundant than VO in solar composition chemical equilibrium), together with the 1000x solar $\rm VO/H_2O$ abundance ratio} needed to fit the spectrum, warrants skepticism for this solution, as pointed out by~\citet{Evans2017}. Moreover, when the retrieved abundances were input into radiative-convective models, no self-consistent solution of the chemical composition and the thermal structure could provide a good fit to the spectrum.
 
In Figure~\ref{fig:Spectrum}, we compare the emission spectrum from our solar composition SPARC/MITgcm simulation to the observations. The model provides a reasonable match for the amplitude of the spectral modulation, which we consider as a success of the model given that no fine tuning has been performed. However, the GCM model is a poor fit to the data with a $\chi^2$ of $89$. This is similar to the best fit blackbody and no worse than the self-consistent radiative convective models presented in~\citet{Evans2017}. Nonetheless, our model provides a more plausible self-consistent solution as it does not require extremely non-solar abundance ratios in the atmosphere of the planet. As an alternative explanation to the highly non-solar abundance ratio, we attribute the lack of strong spectral features in the spectrum to the combined effects of a strong vertical gradient in water abundance due to thermal dissociation and the presence of $\rm H^-$ bound-free absorption at wavelengths bluer than $1.4\mu m$. 

As seen in Fig.~\ref{fig:ContrFunctions}, at the substellar point, the thermal structure is dominated by a very strong positive temperature gradient, ranging from 2500K at 100 mbar to 3500K at 1 mbar. Such a large temperature gradient could result in very large spectral features in the WFC3 bandpass. However, the contribution functions at $1.25\mu m$ and $1.4\mu m$ (shown as solid red and blue lines on the left side of Fig.~\ref{fig:ContrFunctions}) peak almost at the same pressure level; the spectrum probes the same layers throughout the WFC3 bandpass, explaining the lack of strong water emission features. We now investigate the possible reasons for this more thoroughly. 

\subsubsection{The role of water dissociation}

The molecular abundances at the substellar point are also shown in Fig.~\ref{fig:ContrFunctions}. All major molecules, including $\rm H_2$, $\rm H_2O$, TiO and VO but not CO are partly thermally dissociated above the $1.4\mu m$ photosphere. To highlight the role played by molecular dissociation, we calculate a model that uses the same thermal structure as the solar composition global circulation model but assumes a constant water abundance when calculating the spectrum. As shown by the dotted line contribution functions of Fig.~\ref{fig:ContrFunctions}, if water were not dissociated at all and the planet somehow maintained the same thermal structure, the $1.4\mu m$ photosphere would be at a pressure ten times lower than it actually is. Moreover, the pressure of the $1.25\mu m$ photosphere would be five times larger than the pressure of the $1.4\mu m$ photosphere, whereas they are the same in the fiducial model. The consequences on the spectrum can be seen by comparing the red and blue curves of Fig.~\ref{fig:Spectrum}. When thermal dissociation of water is neglected, the layers probed in and out of the water band have a large temperature difference, producing very large spectral features. Although molecular abundances are not strongly depleted at the photosphere itself (given by the maximum of the contribution functions), molecular dissociation creates a vertical gradient in the molecular abundances that plays a fundamental role and explains the shape of the emission spectrum of the planet.

At longer wavelengths, shown in the bottom panel of Fig.~\ref{fig:Spectrum}, our fiducial model exhibits several water emission features. Although molecular dissociation weakens the spectral contrast of the water features, it does not completely erase them. At wavelengths larger than $2\mu m,$ the contrast between the peaks and the troughs of the water cross-section is larger than at shorter wavelengths, and the spectral features become large enough to be observable with instruments such as JWST (see also Fig.~\ref{fig:opacities}). Emission by CO molecules, that are not dissociated at all, also contribute to the spectrum between 2.5 and 3 $\mu m$ and between 4 and 5 $\mu m$. 

\subsubsection{The role of H$^-$}
\begin{figure}[h!]
    \centering
    \includegraphics[width=\linewidth]{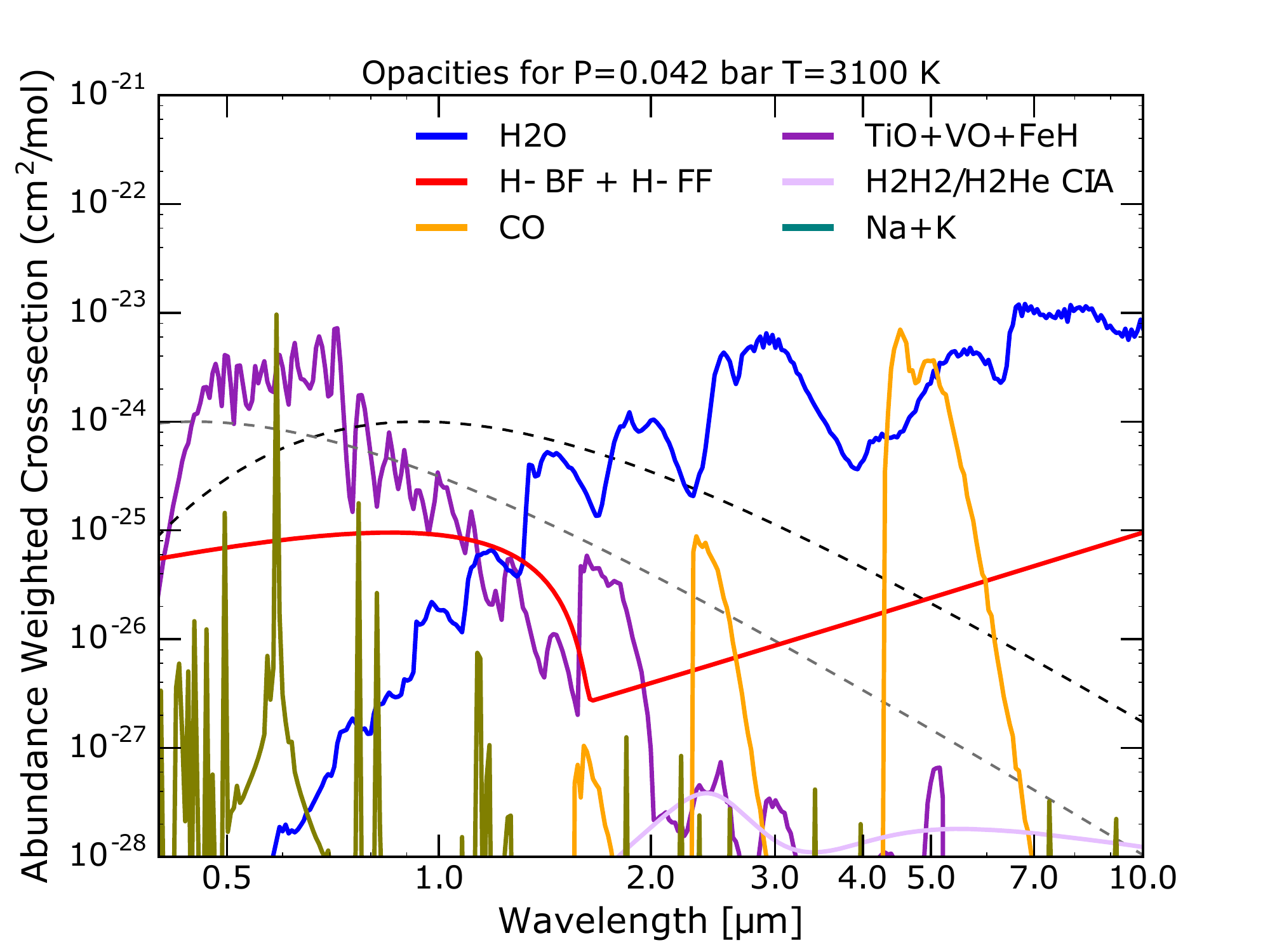}
    \caption{Abundance weighted cross-sections of relevant ions and molecules taken close to the $1.4\mu m$ photosphere of WASP-121b (P=40mbar,T=3100K). The dashed line represents a scaled blackbody emission curve at a temperature of 3100K (black, corresponding to the planet emission) and 6460K (gray, corresponding to the incoming stellar light).}
    \label{fig:opacities}
\end{figure}

At high temperatures, $\rm H_2$ dissociates into $\rm H$, and the ionization of alkali metals Na and K creates a large abundance of free electrons. Electrons and H combine to form a small population of $\rm H^{-}$ ions, with a Volume Mixing Ratio of the order of $10^{-9}$ at the substellar point of WASP-121b. 

$\rm H^-$ can transform into H and e- by absorbing a photon, leading to the bound-free absorption cross-section. H, e- and a photon can also interact together through the free-free cross-section. The bound-free cross-section of $\rm H^{-}$ is quite large and increases with wavelength until a cut-off wavelength of $1.4\mu m$ above which photons do not have enough energy to break the bond between $\rm H$ and its second electron~\citep{Lenzuni1991}. As a consequence, as shown in Fig.~\ref{fig:opacities}, the bound-free absorption opacity is important for wavelengths shorter than $1.4\mu m$ and the free-free opacity is important for wavelengths longer than $1.4\mu m$. 

For $\lambda<1\mu m,$ the bound-free opacity of $\rm H^{-}$ is larger than the sodium opacity outside the narrow core of the line but smaller than the combined opacity from TiO and VO. As a consequence, in a case where TiO and VO are no longer present, $\rm H^{-}$ would become the main absorber of stellar light, thus significantly controlling the atmospheric thermal structure. 

Between $1$ and $1.4\mu m$, both water and TiO/VO opacities are low, leading to a region where $\rm H^{-}$, TiO, and $\rm H_2O$ opacities have similar strength. Particularly, $\rm H^-$ opacity fills the gap between the two water bands at $1.1$ and $1.4\, \mu m$, effectively lowering the contrast between the top and the bottom of the bands. As seen by comparing the plain and dashed lines on the left side of Fig.~\ref{fig:ContrFunctions}, the presence of $\rm H^-$ shifts the  $1.25\mu m$ contribution function toward lower pressures while leaving the $1.4\mu m$ contribution function unchanged. The effect on the spectrum can be seen by comparing the red and orange lines of Fig.~\ref{fig:Spectrum}: the emission feature of $\rm H^-$ between 1 and 1.4$\mu m$ reduces the contrast between the peaks and the troughs of the water band.

As a conclusion, we attribute the lack of strong spectral features in WASP-121b to the combination of two different effects: the presence of a strong gradient of water due to thermal dissociation that weakens the spectral feature of water at $1.4\mu m$ and the presence of $\rm H^-$ bound-free opacities that fills the gap in the water opacity window between $1$ and $1.4\mu m$.

\subsection{Analytic formulation}

We now follow an analytical formalism to understand why, for a given thermal structure, the presence of a molecular gradient weakens the molecular features seen in the spectrum. We assume that the planetary flux at a given wavelength is well represented by the blackbody flux at the temperature where the optical depth reaches a value of two thirds. If the atmosphere becomes optically thick at a pressure $P_{\rm 1}$ at wavelength $\lambda_{\rm 1}$ and at a pressure $P_{\rm 2}$ at wavelength $\lambda_{\rm 2}$, the larger the ratio $P_{\rm 1}/P_{\rm 2}$, the larger the spectral features seen in the planetary spectrum. If $P_{\rm 1}=P_{\rm 2}$ then both wavelengths probe the same atmospheric layer resulting in a blackbody spectrum. 

Assuming that only one molecule contributes to the absorption at wavelength $\lambda$, the optical depth is related to the pressure by the following formula:
\begin{equation}
{\rm d}\tau_{\lambda}=-\frac{\sigma_{i,\lambda}A_i}{\mu g}{\rm d}P
\label{eq:first}
,\end{equation}
where $\sigma_{\lambda}$ is the cross section of the main molecule absorbing at the wavelength $\lambda,$  $A_i$ is its volume mixing ratio, $\mu$ is the mean molecular weight of the atmosphere and $g$ the planet's gravity. We now assume that through the photospheric levels, $A_i$ is a function of pressure with $A_i=A_{\rm 0}\left(\frac{P}{P_0}\right)^{\alpha}$; for $\alpha=0,$ this corresponds to the case with a constant abundance. Inserting the functional form for $A(P)$ and some algebraic manipulation, ~\eqref{eq:first} becomes

\begin{equation}
{\rm d}\tau=-\frac{A_{\rm 0}}{(\alpha+1)P_{\rm 0}^{\alpha}\mu g}\sigma_{i,\lambda}{\rm d}P^{\alpha+1}
.\end{equation} 

In order to isolate the effect of a varying molecular abundance, we now integrate this equation from $\tau=0,P=0$ to $\tau=2/3, P=P_\lambda$ assuming a constant temperature and a constant cross-section and obtain

\begin{equation}
P_{\lambda}=\left(\frac{2}{3}\frac{(\alpha+1)\mu gP_{\rm 0}^\alpha}{A_{\rm 0}\sigma_{i,\lambda}}\right)^{\frac{1}{\alpha+1}}
.\end{equation}

Finally, the ratio of the pressure levels where $\tau=2/3$ at two wavelengths $\lambda_1$ and $\lambda_2$ becomes
\begin{equation}
\left(\frac{P_{\lambda_1}}{P_{\lambda_2}}\right)=\left(\frac{\sigma_{i,\lambda_2}}{\sigma_{i,\lambda_1}}\right)^\frac{1}{\alpha+1}.
\label{eq:final}
\end{equation}
Or, taking the logarithm
\begin{equation}
\log(P_{i,\lambda_1})-\log(P_{i,\lambda_2})=\frac{1}{\alpha+1}(\log(\sigma_{i,\lambda_2})-\log(\sigma_{i,\lambda_1}))
.\end{equation}

It is apparent from equation~\ref{eq:final}, that the larger the dependence of the molecular abundance with pressure, the smaller the ratio between the pressures probed at the two separate wavelengths. In the case of a strong discontinuity, that is, for $\alpha\to\infty$, the pressure levels probed at two different wavelengths are always the same, whatever the spectral structure of the molecular cross-section, and the atmospheric spectrum would tend toward a blackbody spectrum. If $\alpha=0$, corresponding to an atmosphere with constant abundances, the pressure of the photosphere is directly proportional to the cross-sections. 

To first order, $\alpha$ can be estimated using the isothermal limit given in Table~\ref{Table:alpha}. In a real atmosphere, $\alpha$ depends on the actual shape of the pressure-temperature profile. If the temperature decreases with decreasing pressure, $\alpha$ is expected to be smaller than in the isothermal limit whereas if the temperature increases with decreasing pressure (i.e., a thermal inversion), $\alpha$ should be larger than the isothermal limit. 

For water in WASP-121b, based on Fig.~\ref{fig:ContrFunctions}, we find $\alpha\approx2.5$ across the photosphere. This should lead to a pressure difference between the region probed at two different wavelengths dominated by the water absorption, $\log(P_{\rm 1})-\log(P_{\rm 2})$, 3.5 times smaller than in the case of a constant abundance. As shown by the contribution functions of Fig.~\ref{fig:ContrFunctions}, in the case without dissociation and therefore a constant water abundance, $P_{\rm 1.4\mu m}\approx5mbar$ and $P_{\rm 1.25\mu m}=25mbar,$ whereas in the case including water dissociation but neglecting $\rm H^-$ opacities, we obtain  $P_{\rm 1.4\mu m}\approx45mbar$ and $P_{\rm 1.25\mu m}=70mbar$ and we obtain
\begin{equation}
\frac{(\log{P_{1.25\mu m}}-\log{P_{1.4\mu m}})|_{w/o\,\,\,\, diss.}}{(\log{P_{1.25\mu m}}-\log{P_{1.4\mu m}})|_{with \,\,\,\, diss.}}\approx3.5\,,\end{equation}
corresponding to $\alpha+1$ with $\alpha\approx2.5$, as expected.

\subsection{Thermal structure}

The thermal structure of a planet atmosphere is determined by the balanced effect of heating through stellar radiation and cooling through thermal radiation. We can use Fig.~\ref{fig:opacities} to understand which gaseous species will play a role in the radiative balance of the atmosphere. As seen by comparing the blackbody of the star and the opacities, the main absorbers of stellar light are going to be TiO and VO, although $ \rm H^-$ and the thermally broadened alkali lines are also likely to play a role. When comparing the blackbody at the planet temperature to the opacities, we conclude that water, TiO/VO, CO and $\rm H^-$ are going to be the main species responsible for the radiative cooling of the atmosphere. Because CO does not dissociate, it will likely be the only molecule able to radiate away energy at low pressures.

\begin{figure*}
\includegraphics[width=17cm]{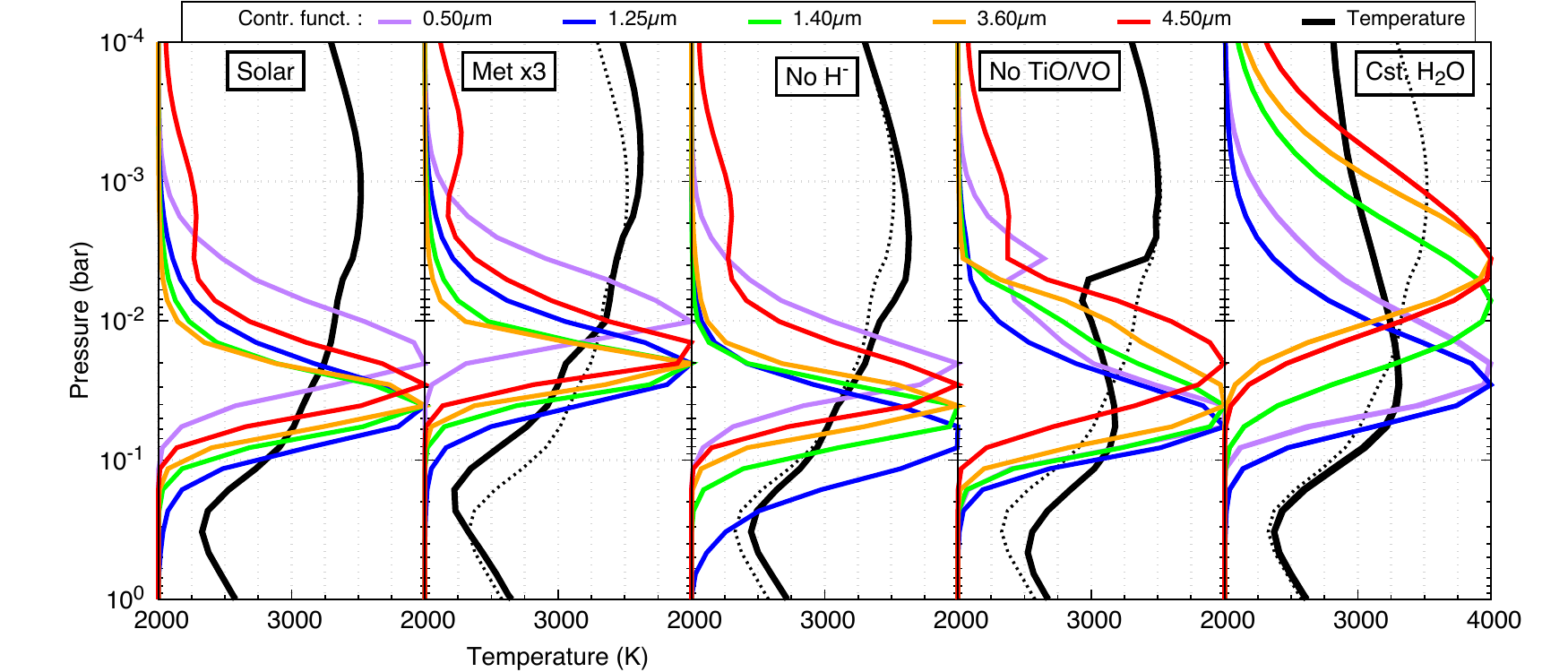}
\caption{Pressure-temperature profile at the substellar point for five different global circulation models of WASP-121b assuming different atmospheric composition: solar, enhanced metallicity (x3), without $\rm H^-$, without TiO/VO and with a constant water abundance. The contribution functions in the optical are also plotted (purple, $0.5\mu m$) and infrared (blue, $1.25\mu m$, green, $1.4\mu m$ and red $4.5\mu m$). The solar composition thermal structure is always shown as a dotted line for comparison.}
\label{fig:ContrAll}
\end{figure*}

\begin{figure*}
\sidecaption
\includegraphics[width=10cm]{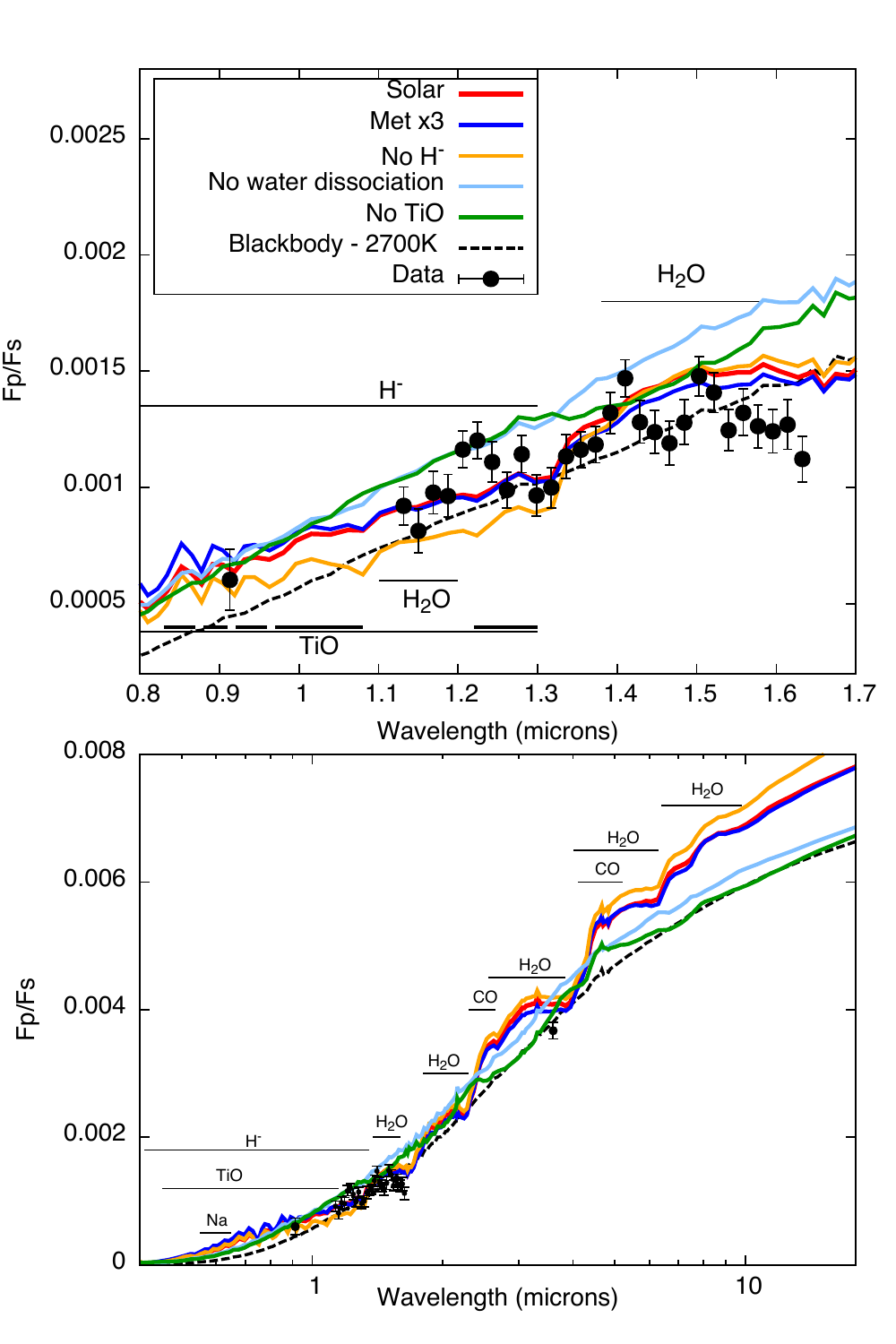}
\caption{Emission spectrum of WASP-121b compared to five different global circulation models of WASP-121b assuming different atmospheric composition: solar, enhanced metallicity (x3), without $\rm H^-$, without TiO/VO and without water dissociation. A model for an isothermal atmosphere at 2700K is shown as a dashed line for comparison. Lines indicating the main spectral bands have been added for clarity. All the models are self-consistently calculating the thermal structure of the planet, contrary to the models of Fig.~\ref{fig:Spectrum} that were post-processed using the thermal structure from the solar composition model.}
\label{fig:SpectrumComp}
\end{figure*}

To understand the radiative balance of the atmosphere more precisely, we now investigate the role of metallicity, $\rm H^-$, TiO/VO and $\rm H_2O$ dissociation on the thermal structure and spectrum of WASP-121b.  We show in Fig.~\ref{fig:ContrAll} the thermal structure and contribution functions for our fiducial model, a metal enriched case, a case without $\rm H^-$, a case without TiO/VO and a case without water dissociation. In terms of energetic balance, the $0.5\mu m$ contribution function indicates approximately at which layers the stellar light is absorbed whereas the contribution function at $1.25\mu m$ and $1.4\mu m$ indicate which layers of the atmosphere are most contributing to the cooling. Comparing the contribution functions at $1.25\mu m$, $1.4\mu m$, $3.6\mu m$ and $4.5\mu m$ provide context to understand the shape of the emission spectrum seen in Fig.~\ref{fig:SpectrumComp}.

\subsubsection{Solar composition case}
In the solar composition case, the contribution function at $0.5~\mu m$ peaks at lower pressures than the contribution functions in the infrared, which explains the presence of a strongly inverted pressure-temperature profile~\citep{Hubeny2003,Fortney2008}. Although both TiO, VO, and water are significantly depleted above the infrared photosphere, the depletion rates of TiO and VO are smaller than the depletion rate of water, meaning that the optical opacities of TiO and VO remain stronger than the infrared opacity of water at all pressure levels. As a consequence, the stellar light is absorbed by TiO and VO at lower pressure than the peak of the planetary emission. As discussed in Sect.~\ref{Sec:EmSpec}, the $1.25\mu m$ and $1.4\mu m$ probe similar depth and the atmospheric features in the WFC3 bandpass are very small. The $4.5\mu m$ contribution function is double peaked, with a maximum at 30mbar and another at around 2mbar (see Fig.~\ref{fig:ContrAll}). The deeper maximum is due to the emission from water whereas the maximum at low pressures is due to the emission from CO molecules. Contrary to water, CO molecules are not dissociated in the atmosphere and probe much lower levels. More generally, as seen in Fig.~\ref{fig:SpectrumComp}, the presence of a thermal inversion can be better diagnosed at wavelengths larger than $2\mu m$ than inside the WFC3 bandpass as both the presence of CO absorption and the larger contrast between the peaks and the trough of the water cross-section allow to probe a larger range of atmospheric pressures~\citep[e.g.,][for WASP-103b]{Kreidberg2018}.

\subsubsection{The role of metallicity}

We increase the metallicity in our model by multiplying all heavy-element abundances by a factor 3. To first order, this should increase all molecular abundances and therefore opacities by a factor 3 and therefore shift the pressure temperature profile by a factor 3 in pressure. As seen in Figure~\ref{fig:ContrAll}, the temperature pressure profile is indeed shifted upward in the increased metallicity case. However, the shift corresponds to a factor $2$ in pressure, with the $1.4\mu m$ photosphere changing from $40\,\rm mbar$ in the solar composition case to $20\,\rm mbar$ in the three times metallicity case. The smaller-than-expected upward shift can be explained by the temperature dependence of the thermal dissociation. For the higher-metallicity case, the pressure-temperature profile has a stronger thermal inversion, and therefore the water abundance decreases faster with pressure, partially compensating the expected upward shift of the photosphere.

\begin{figure}[h]
\includegraphics[width=\linewidth]{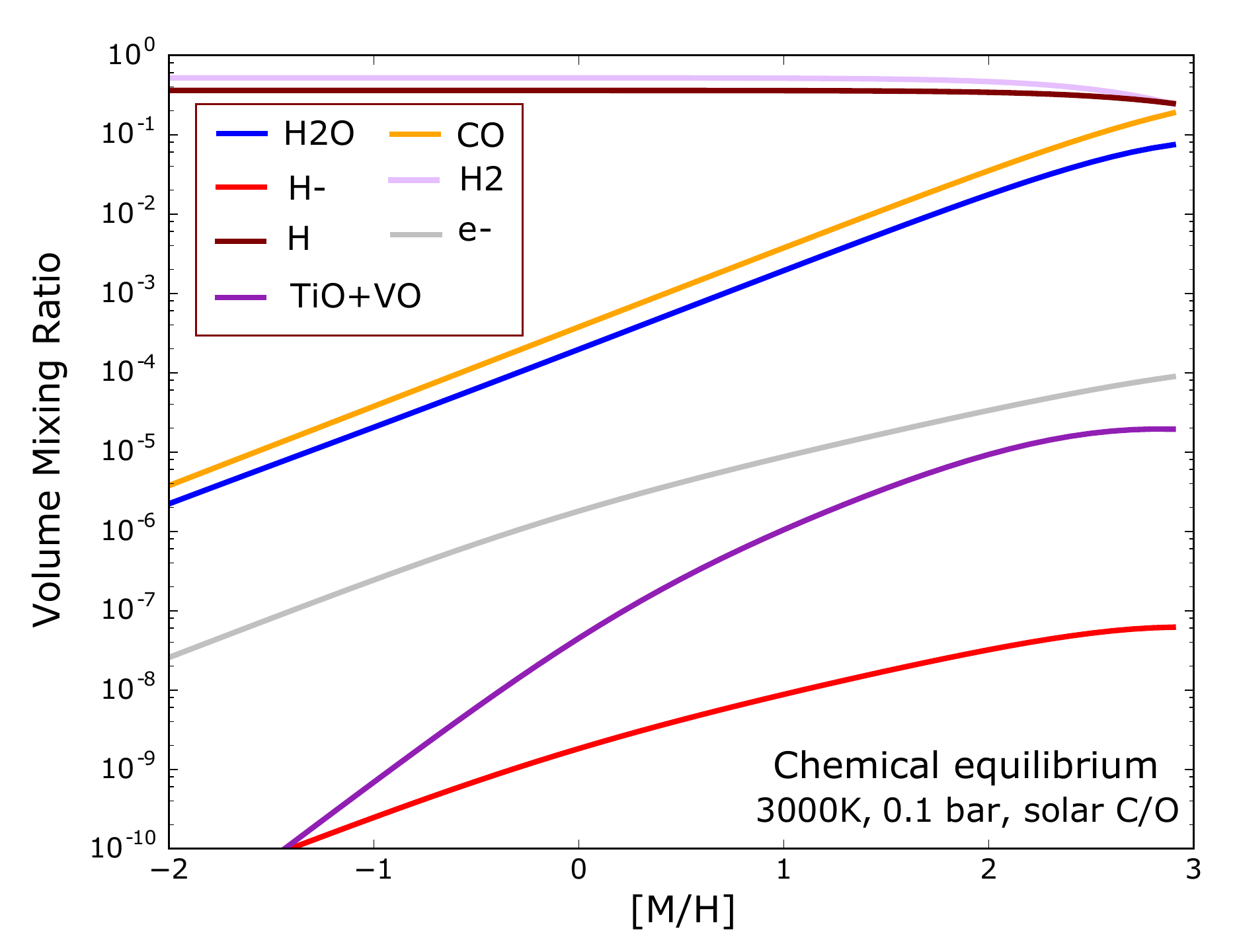}
\caption{Volume mixing ratio of different molecules and atoms as a function of metallicity calculated using the NASA CEA Gibbs minimization code (see Sect.~\ref{sec:Abundances}).The mixing ratios of TiO and VO do not vary linearly with metallicity leading to a change with metallicity of the relative abundance of TiO and VO compared to water.\vspace{1.5cm}}

\label{fig:AbunMet}
\end{figure}

The larger thermal inversion is mainly due to an increase by a factor $\approx 2$ of the photospheric TiO/$\rm H_2O$ abundance ratio between the solar and the three times solar metallicity. This change is due to the combination of two distinct effects. First, contrary to the water abundance, the TiO abundance does not scale linearly with metallicity. The ratio between the TiO abundance and the $\rm H_2O$ abundance at a given pressure-temperature point is increased by a factor $\approx 1.5$ when increasing the metallicity from solar to three times solar (see Fig.~\ref{fig:AbunMet}). Second, due to molecular dissociation, the $\rm TiO/H_2O$ abundance ratio should vary with the square root of pressure at a given metallicity (see Table~\ref{Table:alpha}) and therefore by a factor $1.5$ between the solar and the three times solar metallicity photosphere.

As seen in Fig.~\ref{fig:SpectrumComp}, the overall effect of the metallicity on the spectrum is subtle. As metallicity increases, the thermal inversion becomes stronger and the emission features of CO and TiO become stronger, while the spectral features of water become shallower.

\subsubsection{The role of H$^-$}
 In our fiducial model, as temperature increases and the pressure decreases, the relative abundance of H$^-$ compared to $\rm TiO$ and $\rm H_2O$ increases. As such, in the 1-10mbar region H$^-$ opacities play a significant role in the cooling properties of the atmosphere. Removing $\rm H^-$ decreases the ability of the atmosphere to cool down through radiation in the 1-10mbar region, leading to hotter atmospheric layers. Indeed, when $\rm H^-$ opacities are neglected, the pressure temperature profile becomes hotter in the 1-10mbar region (see Fig.~\ref{fig:ContrAll}).

When looking at the modeled spectrum without $\rm H^-$ (orange model) of Fig.~\ref{fig:SpectrumComp}, the planet is brighter at wavelengths longer than $1.4\mu m$ but dimmer at wavelengths shorter than $1.4\mu m$. This change is due to the larger spread of the contribution functions. Even though the temperature profile is hotter at all pressure levels, the photosphere between 1 and 1.4$\mu m$ is deeper in the case without $\rm H^-$. As a consequence, the mean temperature at the $1-1.4\mu m$ photosphere is smaller in the case without $\rm H^-$ than in the case with $\rm H^-$. This leads to a larger contrast between the water band and the continuum (i.e., between $1.4\mu m$ and $1.25\mu m$). 

\subsubsection{The role of TiO/VO}
We now assume that TiO and VO do not contribute to the overall opacities.  It has previously been proposed that TiO and VO molecules could rain out of the atmosphere because of their condensation on the nightside of the planet~\citep{Parmentier2013}. This mechanism was suggested by~\citet{Beatty2017a} as an explanation for the non-inverted thermal structure they inferred from the HST/WFC3 observations of Kepler-13Ab. 

When TiO and VO are removed from the atmosphere, $\rm H^-$ becomes the main absorber of stellar light. Given that the optical opacity of $\rm H^-$ is not larger than the thermal opacity of water, the stellar light is absorbed close to the layers from which the atmosphere can efficiently cool down to space. As a consequence, the temperature decreases with decreasing pressure from 100 mbar to 5 mbar. Between 5 and 3 mbar, the sign of the temperature gradient changes and the temperature increases with decreasing pressure. This is due to localized heating by sodium, as revealed by the presence of second maximum in the $0.5\mu m$ contribution function. Although sodium is significantly depleted by thermal ionization, the sodium abundance does not drop as fast with decreasing pressure as the molecular abundances (see Figs.~\ref{fig:Abundances} and~\ref{fig:ContrFunctions}). Given the lack of efficient cooling by the molecular opacities, the atmosphere needs to be hotter to re-emit the energy absorbed by the sodium atoms and the temperature is significantly increased in the 5 and 3 mbar region. At lower pressures, the temperature profile is again non-inverted. 

\begin{figure}[h!]
\includegraphics[width=\linewidth]{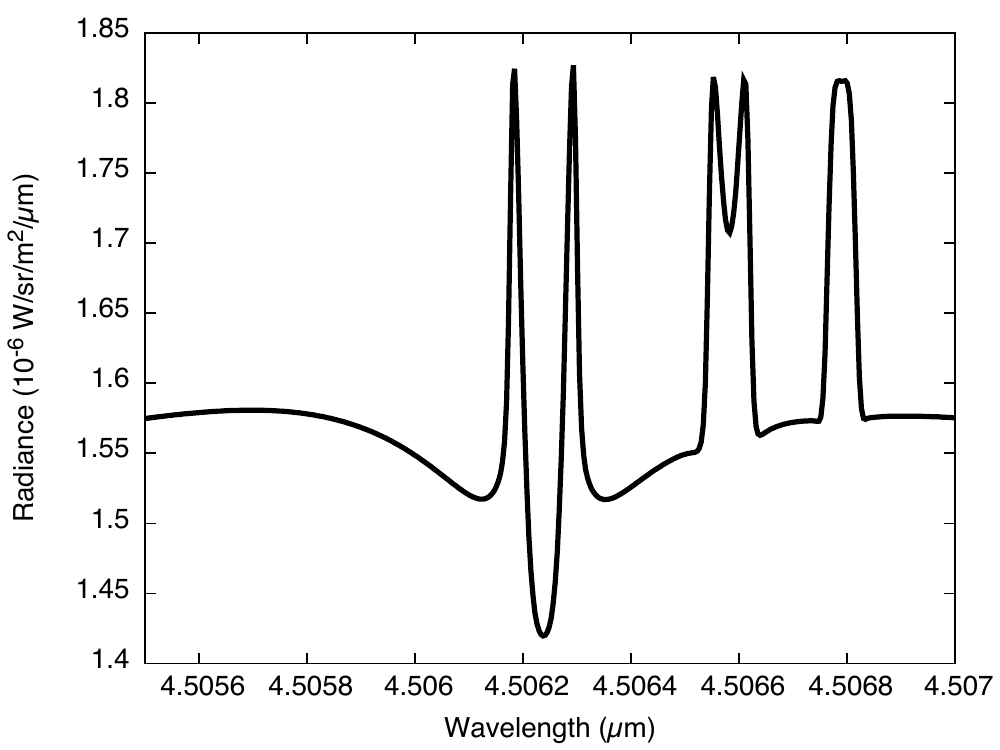}
\caption{Emission spectrum of WASP-121b at high resolution in the case without TiO/VO. Only CO was taken into account in the calculation. The spectrum was calculated with the Planet Spectrum Generator~\citep{Villanueva2018} using the substellar point temperature and CO abundance calculated by the GCM. Three lines of different strength (increasing from left to right) show different shapes as they probe different parts of the temperature-pressure profile.}
\label{fig:COLine}
\end{figure}

While the water bands all probe the non-inverted layers, the CO band at $4.5\mu m$ probes above the thermal inversion. As seen in Fig.~\ref{fig:SpectrumComp}, this produces both broadband absorption features due to water (e.g., at $1.4\mu m$ or $3.6\mu m$) and broadband emission feature due to CO (e.g., at $4.5\mu m$) in the same spectrum. When examined at high resolution (see Fig.~\ref{fig:COLine}), the shape of the CO lines is quite complex. For the strongest lines, probing a wide range of pressures, the core and the wings of the line are seen in absorption but the core is offset from the wings of the line due to the thermal inversion. This would introduce further complexity when trying to cross-correlate the high-resolution spectrum of a planet with a template spectrum~\citep[e.g.,][]{Brogi2014}. Specifically, a technique more sensitive to the core of the line would conclude toward a non-inverted atmosphere whereas a technique more sensitive to the broad shape of the line would conclude toward an inverted thermal structure.

\subsubsection{The role of water dissociation}
We now assume that water does not dissociate, that is, that water has a constant abundance throughout the atmosphere while other molecules such as TiO and $\rm H_2$ are still being thermally dissociated. This is not a realistic case as it is unlikely that a mechanism exists to inhibit the thermal dissociation of water and not the dissociation of other molecules. It is, however, an interesting experiment to understand the role of the radiative cooling of water molecules. As shown in the right panel of Fig~\ref{fig:ContrAll}, the $0.5\mu m$ contribution function is approximately similar to the case with water dissociation as TiO and VO abundances are similar in both cases. The infrared photospheres, however, are shifted to much lower pressures. As a consequence, the stellar light is absorbed at deeper pressures than the planet radiates and the pressure temperature profile decreases with altitude. Moreover, the contribution functions probe very different depths with wavelengths. Despite this fact, the spectrum shown in Fig.~\ref{fig:SpectrumComp} shows only extremely shallow absorption features and exhibits a blackbody shape. Contrary to the other cases, the pressure-temperature profile at the substellar point is not a good proxy for the dayside thermal structure as a large area around the substellar point shows a much shallower temperature gradient at low pressures, leading to extremely damped features.

\section{3D structure}
\label{Sec:3DStruct}
\subsection{Horizontal thermal and chemical structure}
\label{Sec:Phasecurve}
Highly irradiated tidally locked planets have large atmospheric temperature contrasts because they re-emit the energy received on the dayside faster than winds can transport it to the nightside. This large planetary-scale temperature contrast can lead to strong horizontal variations in the chemical composition of the planet.

\begin{figure*}
\sidecaption
\includegraphics[width=10cm]{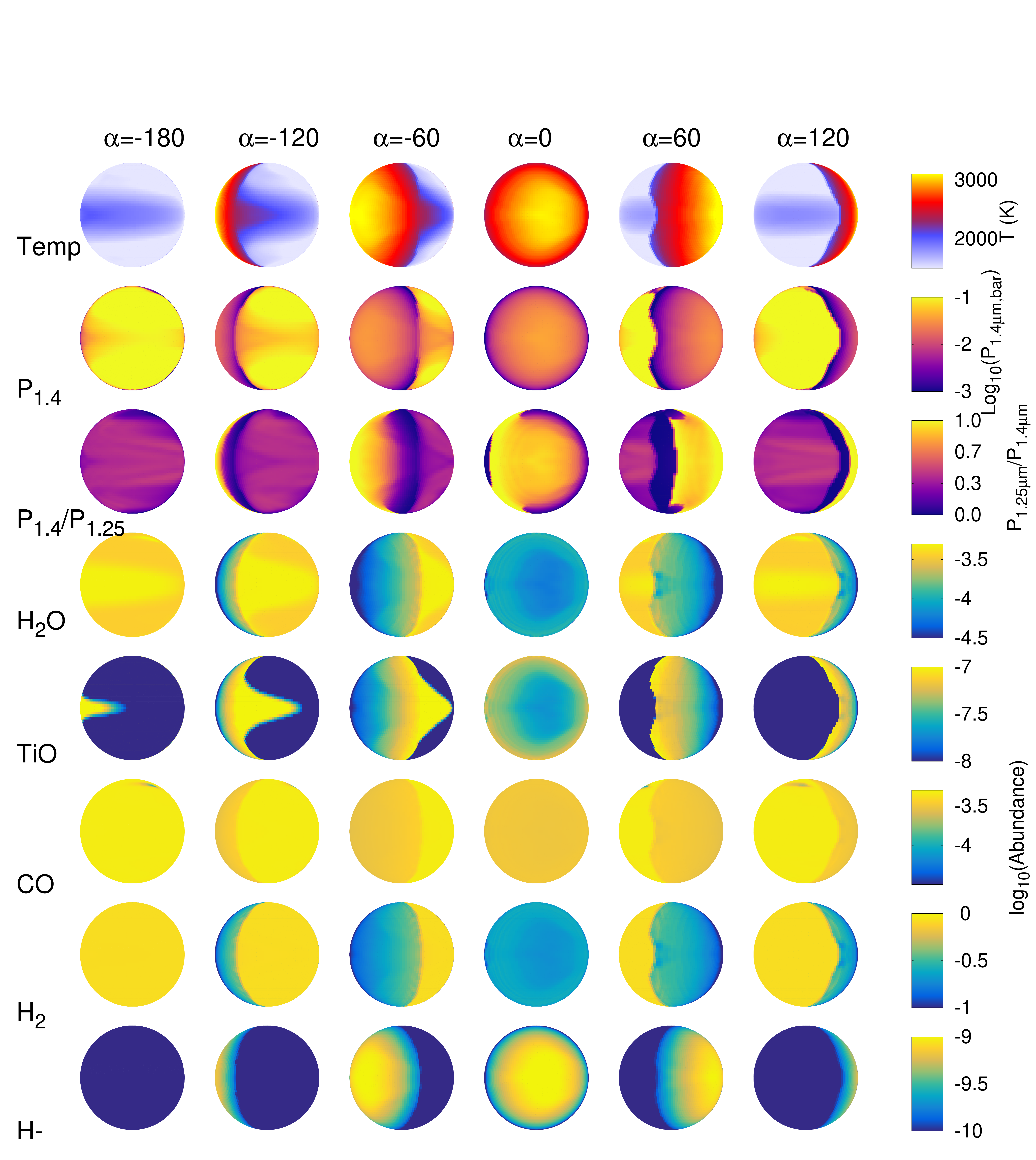}
\caption{Structure of the visible hemisphere of WASP-121b as a  function of phase from the solar composition SPARC/MITgcm model. $\alpha=0$ is the dayside as seen during secondary eclipse and $\alpha=-180$ is the nightside as seen during transit. The first two rows show temperature and pressure at the $1.4\mu m$ photosphere. The third row shows the ratio of the photospheric pressures at $1.4\mu m$ and $1.25\mu m$, a proxy for the amplitude of the water feature in the WFC3 bandpass. The last five rows show the abundances of important molecules slightly above the $1.4\mu m$ photosphere, defined where the contribution function reaches three quarters of its maximum value.}
\label{fig:map}
\end{figure*}

Figure.~\ref{fig:map} shows the photospheric properties of the planet hemisphere visible from Earth at different orbit phases as calculated with our fiducial, solar-composition model.

The dayside photosphere is hot and varies from 2500K at the limb to 3000K at the substellar point. Most molecules are dissociated on the whole dayside, including $\rm H_2$, $\rm H_2O,$ and $\rm TiO,$ while $\rm H^-$ is present. The photospheric regions between the substellar point and +/-60$^{\circ}$ around the substellar point are relatively homogeneous with a quasi-constant photospheric water abundance. This is because the vertical water gradient sets the photospheric level at $1.4\mu m$ relatively independently of the viewing angle. As we go away from the substellar point, the pressure-temperature profiles become cooler, leading to a larger abundance of water at a given pressure. In turn, this moves the photosphere to lower pressures, until water becomes too dissociated to be an important opacity source. 

As we go from the center to the limb of the dayside, the temperature drops at a given pressure level. However, the further from the substellar point, the lower the pressure of the photosphere. Given that the temperature increases with decreasing pressure, this leads to a quasi-constant photospheric temperature around the substellar point. As a consequence, the hot spot of the planet appears more diluted in the photospheric maps than in the isobaric map, { which should lead to an increase in the width of the phase curve around the maximum}. Close to the limb, the effect of the viewing angle effect dominates again, the photospheric pressure becomes very small, and the water abundance at the photosphere becomes smaller. 

The nightside photosphere is $\approx1200\rm\, K$ cooler than the dayside photosphere, the molecules are not dissociated, $\rm H^-$ is not present and TiO is condensed out. Although there is more water available, the cooler temperatures lead to a more compact atmosphere and therefore a larger photospheric pressure at $1.4\mu m$. 

At the limb, the photospheric temperature changes by 1000K in a few degrees of longitude. Numerous physical processes are expected to take place at the limb, including the recombination of $\rm H_2$, which should release a significant amount of latent heat \cite{Bell2018} (not taken into account in our simulations), the recombination of $\rm H_2O$, the recombination and condensation of molecules such as $\rm TiO,$ and more generally the formation of condensates such as iron, $\rm CaTiO_3$, $\rm Al_2O_3$, $\rm MgSiO_3$ and so on~\citep{Parmentier2016}. The interplay between the transport timescale, the recombination timescale, and the cloud formation timescale are going to be crucial to understand the properties of the atmospheric limb. As seen in section~\ref{sec:Transmission}, both condensed and gaseous $\rm TiO$ seem important to understand the transmission spectrum of WASP-121b, pointing toward the idea that molecules have the time to both recombine and condense when transported from the substellar point to the atmospheric limb. 

The spectral contrast within the WFC3 bandpass can be estimated using the third row of Fig.~\ref{fig:map}. When the photospheric pressures at $1.4\mu m$ and $1.25\mu m$ are equal, as is the case in most of the dayside, only small features are present in the WFC3 spectrum. In the nightside, the $1.25\mu m$ wavelength should probe pressures three times larger than the $1.4\mu m$ wavelength and strong molecular features are expected.

\subsection{Phase curve}
\begin{figure}[h]
\includegraphics[width=\linewidth]{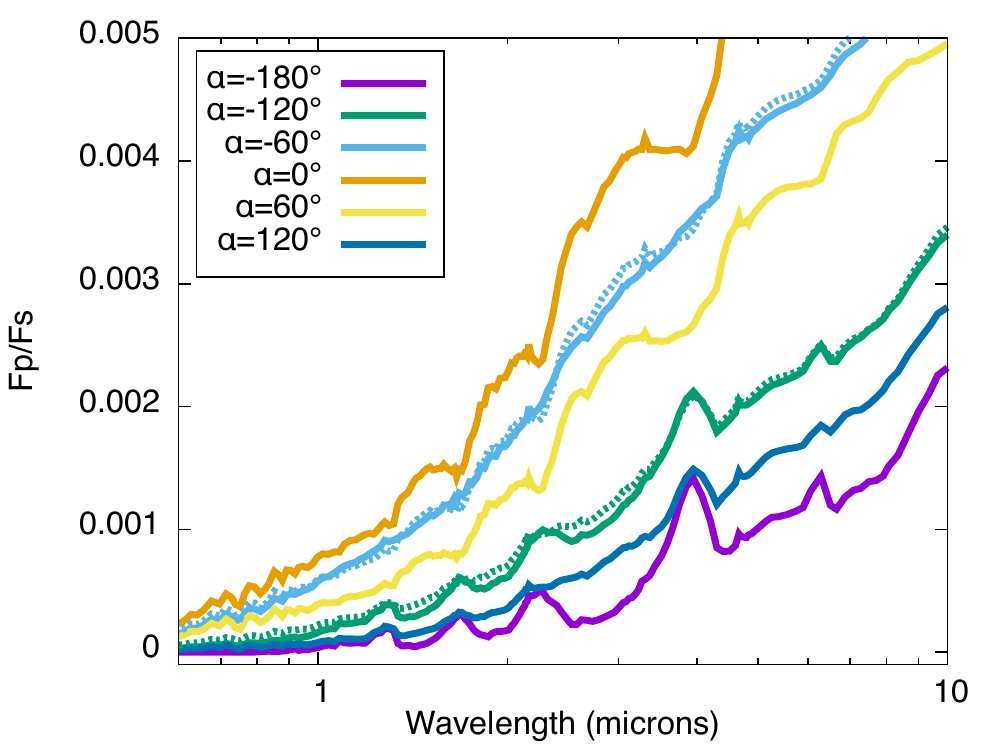}
\caption{Spectrum of WASP-121b at different orbital phases, $\alpha=0^{\circ}$ being the secondary eclipse. Emission features are expected for $-120^{\circ}<\alpha<60^{\circ}$ whereas absorption features are seen for $\alpha<-120^{\circ}$ and  $\alpha>60^{\circ}$. The spectra at phases $-120^{\circ}$ and $-60^{\circ}$ \textcolor[rgb]{0.984314,0.00784314,0.027451}{are} well represented by an area weighted sum of the dayside and the nightside spectrum (dotted line) but the $120^{\circ}$ and $60^{\circ}$ spectra are not.}
\label{fig:PC}
\end{figure}

Phase curve observations can provide the hemispherically averaged spectrum of the planet at different orbital phases and reveal its 3D structure~\citep[see][for a review]{Parmentier2017}. As seen in Fig.~\ref{fig:PC}, the spectrum of the visible hemisphere of the planet shows emission features at phases $0^{\circ}$ and $60^{\circ}$, a featureless spectrum at phases $-60^{\circ}$ and $120^{\circ}$, and absorption features at phases $-120^{\circ}$ and $-180^{\circ}$. Also shown in Fig.~\ref{fig:PC} is a linear combination of the dayside and nightside spectra where each spectrum was weighted by the dayside or nightside projected area of the planet seen at a given phase. The near-blackbody spectrum seen at $-60^{\circ}$ can be explained as the combination of the emission bands from the dayside and the absorption bands of the nightside. The spectrum at $-120^{\circ}$ can also be well explained as the sum of the nightside and the dayside spectrum. The spectra at $60^{\circ}$ and $120^{\circ}$ are not, however, well represented by a combination of the mean dayside and the mean nightside spectrum. This is due to the fact that both the dayside and the nightside are not homogeneous. At phases $60$ and $120,$ the planet shows the coldest point of the nightside and the coldest point of the dayside, which are not dominant in the dayside and nightside averaged spectrum. At phases $-60^{\circ}$ and $-120^{\circ}$ the planet shows the hottest points of the dayside and the nightside which are more representative of the mean dayside and nightside spectrum.

\subsection{Clouds and transmission spectrum}
\label{sec:Transmission}
\begin{figure}[h!]
\hbox{\hspace{3em} \includegraphics[width=0.9\linewidth]{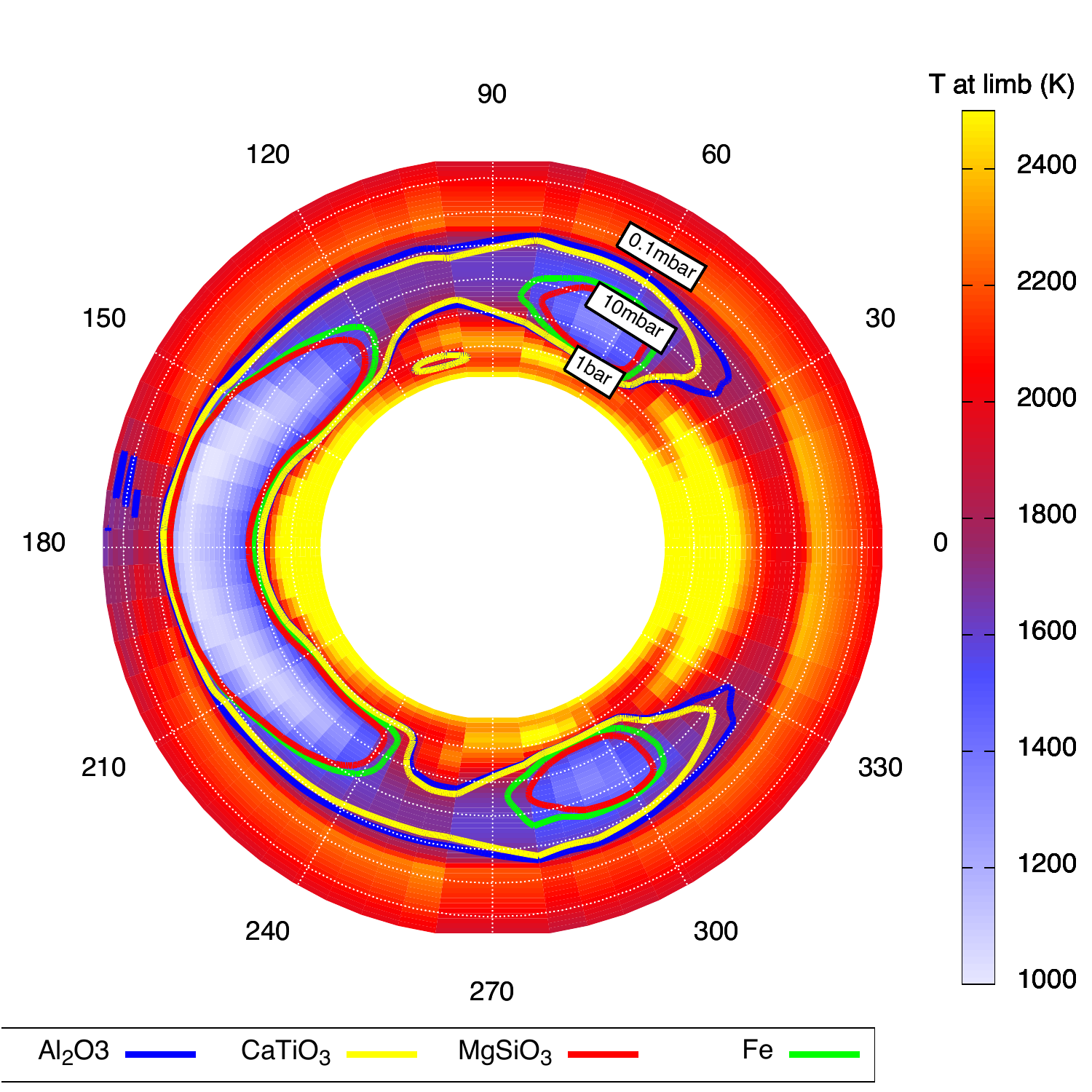}}\includegraphics[width=\linewidth]{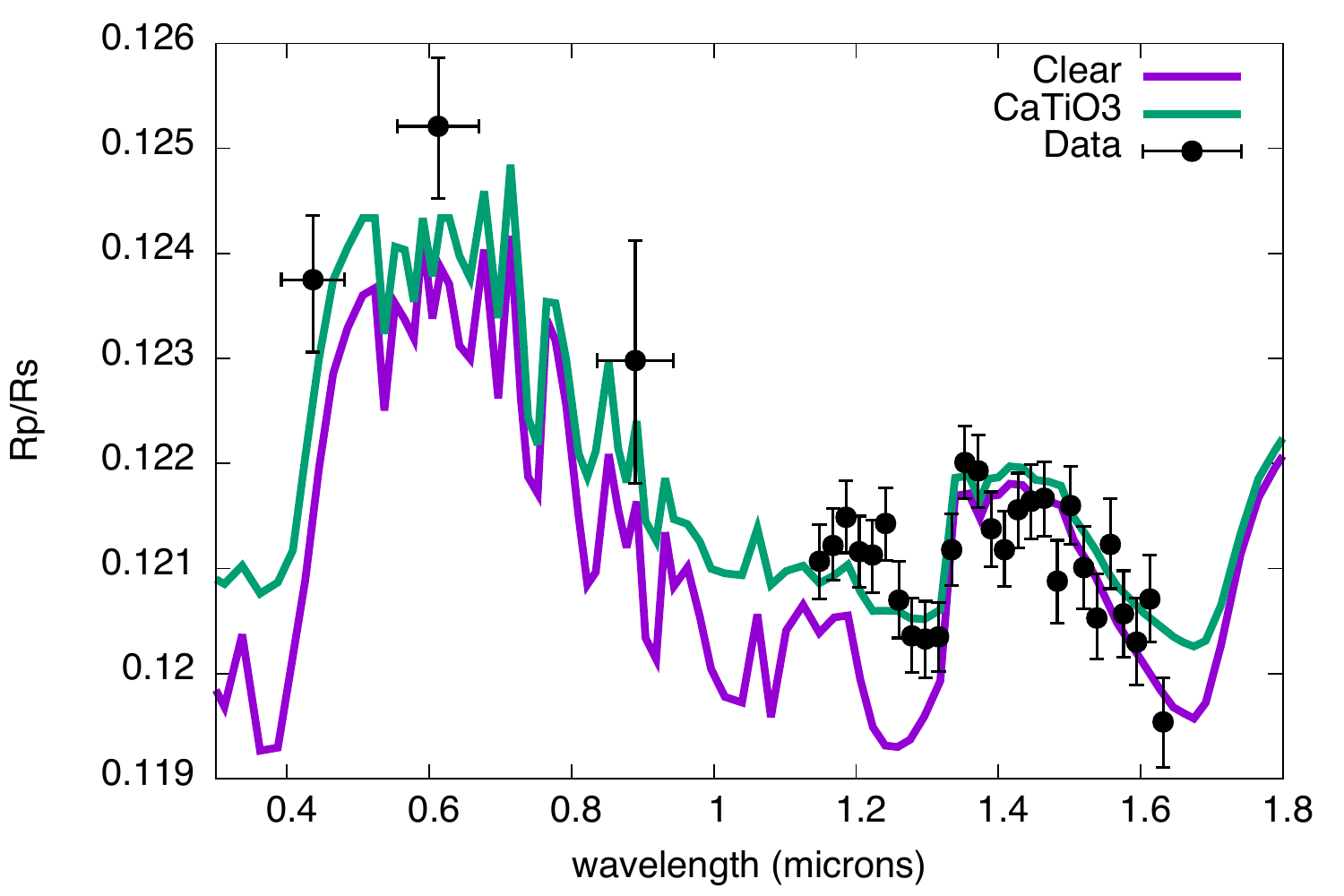}
\caption{\emph{Top} Temperature map at the limb of WASP-121b as predicted by a solar composition SPARC/MITgcm. The east limb is at an azimuth of $0^{\circ}$. The condensation curves of several possible condensates have been overplotted. The atmosphere should be cloudy where it is cooler than the condensation curve. \emph{Bottom} Transmission spectrum of WASP-121b corresponding to the above temperature map with and without taking $\rm CaTiO_3$ clouds into account. We assume a particle size of $0.1\mu m$, that the particles neither settle nor are transported by the circulation, and that the number of particles is given by the mass conservation of condensed Ti atoms.  The models are compared to  data taken from~\citet{Evans2016}.}
\label{fig:Transmission}
\end{figure}

Figure~\ref{fig:Transmission} shows the temperature at the limb of WASP-121 as predicted by solar composition SPARC/MITgcm simulation. The temperature map is highly asymmetric, with a west limb cooler by 1000 K than the east limb at pressures ranging from 1bar to 0.1mbar. The east/west limb temperature contrast is driven by the presence of a super rotating equatorial jet that transports heat from the substellar point to the east limb and from the anti-stellar point to the west limb. At pressures  greater than 1bar, the radiative timescale is larger than the advective one and the atmospheric circulation can better homogenize the temperature. At pressures lower than 0.1mbar, the radiative timescale is short compared to the advective one and the atmospheric circulation is characterized by an axisymmetric substellar to anti-stellar circulation pattern~\citep{Showman2008a}. As a consequence, the temperature is quite symmetric in the upper layers of the atmosphere. Importantly, the temperatures at the limb are too cold to allow the thermal dissociation of water or TiO. If the recombination timescale is short enough compared to the transport timescale of a few hours (see \S\ref{Sec:Phasecurve} above), then molecules should be present at the limb of WASP-121b.
Clouds can possibly form in part of the limb of WASP-121b. Between 1bar and 0.1bar and between $60$ and $300$ degrees of azimuth the temperatures are cold enough for corundum ($\rm Al_2O_3$) and perovskite ($\rm CaTiO_3$) condensates to form. Iron and silicates should also form but cover a smaller area of the limb.

In the bottom panel of Fig.~\ref{fig:Transmission}, we compare the transmission spectrum of WASP-121b obtained by~\citet{Evans2016} with our model. Although our cloudless model is consistent with the optical points, it overestimates the amplitude of the water absorption feature at $1.4\mu m$. To explain this discrepancy,~\citet{Evans2016} proposed that a supersolar abundance ratio of FeH over water could increase the opacities around $1.2\mu m$ and improve the fit to the observations. Here we propose an alternative explanation: as shown by the green curve, when the presence of $\rm CaTiO_3$ cloud opacity is included in the model, the amplitude of the water feature is reduced while the shape of the water feature is kept similar, leading to a much better fit of the spectra. This is a direct consequence of the presence of a partially cloudy atmosphere~\citep{Line2016,Kempton2017}. Interestingly, since TiO is still present as a gaseous species above the 0.1mbar cloud top, and the spectral band of TiO is still apparent in the transmission spectrum and is not strongly affected by the presence of condensed $\rm CaTiO_3$ below it. The fact that both $\rm CaTiO_3$ clouds and gaseous TiO are needed to explain the transmission spectrum of WASP-121 points towards the lack of a cold-trap of gaseous TiO by the clouds~\citep{Parmentier2013,Spiegel2009} and is a hint that the vertical mixing in WASP-121b is strong above the 1 bar level.

\subsection{Recombination timescales}

{ Although recombination of H$_{2}$ is extremely fast under these conditions \citep[e.g.,][]{Bell2018}, no data are available for direct estimates of TiO recombination timescales.  Experimental measurements show fast kinetics for Ti + O$_{2}$ $\rightarrow$ TiO + O at oxidizing conditions \citep[NIST Kinetics Database; cf.][for X + O$_2$ $\rightarrow$ XO + O timescales in silicate vapor atmospheres]{Fegley2014}.  However, this oxidation pathway is unlikely to play a significant role in H$_{2}$-rich atmospheres where molecular oxygen is scarce.  Instead, the recombination of metal oxides may plausibly proceed via a number of alternative reaction pathways, including:
\begin{subequations}
\begin{align}
\textrm{Ti} + \textrm{OH} & \rightarrow  \textrm{TiO} +\textrm{H} \label{rxn_1}\\ 
\textrm{O} + \textrm{H}_{2} & \rightarrow  \textrm{OH} +\textrm{H} \label{rxn_2}\\ 
\textrm{H} + \textrm{H} + \textrm{M} & \rightarrow  \textrm{H}_{2} +\textrm{M}\label{rxn_3}\\[-9pt]
\cline{1-2}
\textrm{Ti} + \textrm{O} + \textrm{M} & \rightarrow  \textrm{TiO} +\textrm{M,} \tag{9, net}
\end{align}
\end{subequations}
where M is any third body.  In this mechanism, reactions \ref{rxn_2} and \ref{rxn_3} (i.e., H$_{2}$ recombination) proceed extremely rapidly, so reaction \ref{rxn_1} is the rate-limiting step with a chemical lifetime given by $\tau_{\textrm{Ti}} = 1/(k_{\ref{rxn_1}}[\textrm{OH}])$, with $[\textrm{OH}]$ the abundance of OH in chemical equilibrium. No rate constant measurements are available for the gas-phase reaction \ref{rxn_1}.  However, rate constants for analogous oxidation reactions of metals (e.g., Fe+OH$\rightarrow$FeO+H in~\citet{Decin2018} or other similar reactions in~\citet{Mick1994}) yield chemical lifetimes on the order of $\lesssim$0.1 s at 0.1bar and 2000 K.  Recombination of TiO is therefore expected to occur on much shorter timescales than the advection timescales. For water, based on reactions R15 and R21 of~\citet{Decin2018}, we expect the recombination to be even faster.}

\section{The family of ultra hot Jupiters}
\label{sec:pop}
\subsection{Comparing four ultra hot Jupiters}
We now use the SPARC/MITgcm global circulation model to model the atmospheres of four different hot Jupiters for which secondary eclipses exist in the WFC3 and Spitzer bandpasses: WASP-121b ($g=8.5m/s^{2}$, $T_{\rm day}\approx2800\,\rm K$), WASP-103b ($g=16m/s^{2}$, $T_{\rm day}\approx2960\,\rm K$), HAT-P-7b ($g=22m/s^{2}$, $T_{\rm day}\approx2640\,\rm K$) and WASP-18b ($g=190m/s^{2}$, $T_{\rm day}\approx2870\,\rm K$). More information on the last three models can be found in~\citet{Kreidberg2018},~\citet{Mansfield2018}, and~\citet{Arcangeli2018}, respectively.

\begin{figure}[h!]
\includegraphics[width=\linewidth]{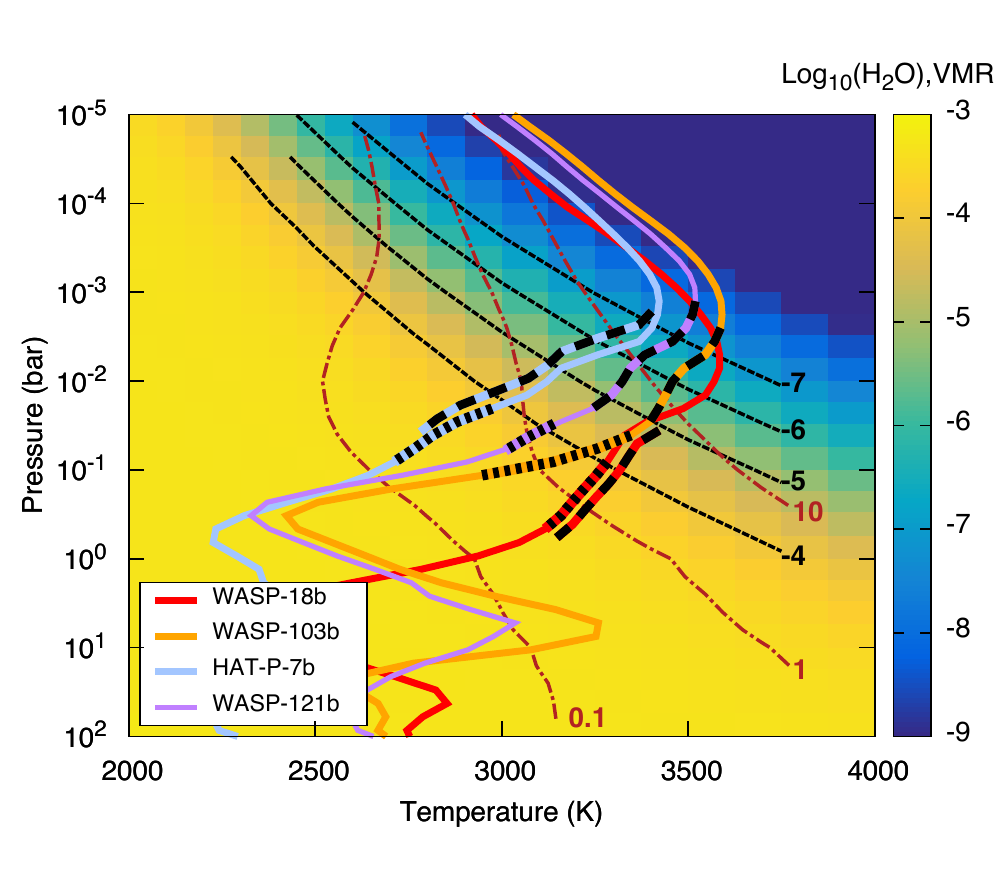}
\caption{Pressure temperature profiles at the substellar point of the four planets studied here. The short dash portion of the profiles are the $1.4\mu m$ photospheric pressures. The long-dash portions represent what the photospheric pressures would be if water dissociation were not taken into account. The background and the black contour lines show the water abundance for a solar composition gas as a function of pressure and temperature. The dashed-dot brown contour lines are the contour lines of the ratio between the opacities of molecules (mainly H$\rm_2$O and TiO) and the bound-free opacity of $\rm H^{-}$ calculated at $1.25\mu m$. At higher temperatures and lower pressures than the line labeled "1", $\rm H^-$ is an important opacity source in the atmosphere.}
\label{fig:PTprofile}
\end{figure}

We show in Figure~\ref{fig:PTprofile} the pressure-temperature profiles at the substellar point obtained from the GCM calculations for the four different planets. On the background, we show the water abundance, modulated mainly by the thermal dissociation of water at high temperature and low pressure. All the pressure-temperature profiles have a similar shape but are shifted toward higher temperature for planets with a larger equilibrium temperature (e.g., HAT-P-7b vs. WASP-103b) and toward a deeper pressure for planets with a higher gravity (e.g., WASP-18b vs. WASP-121b). 

{ The short dash portions of the pressure temperature profiles shows the photosphere of the planet at $1.4\mu m$, that is, inside the water absorption band}. For HAT-P-7b and WASP-18b, the volume mixing ratio of the water at the photosphere of the planets is close to the value of $10^{-3.3}$, expected when the dissociation of water is not taken into account. For the hot and low gravity planets WASP-103b and WASP-121b, however, the water volume mixing ratio at the $1.4\mu m$ photosphere is in between $10^{-4.5}$ and $10^{-3.3}$, that is, the water abundance varies by a factor of ten between the lowest and highest pressures probed by the WFC3 instrument. All the pressure-temperature profiles have a strong thermal inversion at the photosphere, due to the combination of absorption by TiO~\citep{Fortney2008} and the lack of water as a coolant at low pressures~\citep{Molliere2015}. The pressure-temperature profile calculated by the GCM at pressures larger than 1 bar is in a transient state due the large radiative timescale at these high pressures. The solution below 1 bar is therefore highly dependent on the initial condition~\citep{Amundsen2014}, and should not be taken into account. As shown in~\citet{Showman2009}, longer integration does not significantly change  the photospheric temperatures and therefore the emission spectrum of the planet. 

{ The long-dash portions of the pressure temperature profiles show where the photosphere of the planet would be if water did not dissociate, an assumption often made when interpreting the spectra of these hot planets~\citep{Sheppard2017,Haynes2015,Evans2017}. In all four planets, dissociation of water reduces the range of pressures probed by the WFC3 instrument and increases the pressure probed by the observations.} 

Figure~\ref{fig:PTprofile} also shows as dot-dashed lines the ratio between the $\rm H^-$ opacities and all the other molecular opacities at $1.25\mu m$. If this ratio is close to or larger than 1 at the photosphere of the planet, $\rm H^-$ opacities need to be taken into account to correctly interpret the emission spectrum of the planet in the $1-1.4\mu m$ range. $\rm H^-$ opacities should therefore be important in the spectrum of WASP-121, WASP-18b, and WASP-103b, but probably less important to interpret the spectrum of HAT-P-7b or cooler objects. 

\subsection{A population-wide view}

We now address for which planets the spectral properties of the atmosphere are going to be shaped by the thermal dissociation of water molecules. For this we ask whether a self-consistent solution with a constant water abundance is realistic for any given planet. 

{ In the case of a constant water abundance, the $1.4\mu m$ photosphere of a planet can be calculated by integrating equation~\ref{eq:first} assuming that only water absorbs at this wavelength, that the mean molecular weight, gravity, and water abundance are constant with height. We further assume that the cross-section of water is constant with both pressure and temperature at $1.4\mu m$ leading to the simple formula:
\begin{equation}
P_{1.4\mu m}=\frac{2}{3}\frac{g\mu}{\sigma_{\rm H_2O,1.4\mu m}*A_{\rm H_2O}}.
\label{eq:third}
\end{equation}
We fix the volume mixing ratio of water to its abundance at 10 bar, $A_{\rm H_2O}=10^{-3.3}$, corresponding to a solar abundance of water when thermal dissociation is not affecting the abundances. We fix $\mu=2.2$ corresponding to a solar-composition atmosphere where $\rm H_2$ does not dissociate.  $\sigma_{\rm H_2O,1.4\mu m}$ is the absorption cross-section of water at $1.4\mu m$ and approximately equal to $10^{-21} cm^2/molecule$, a value roughly constant with pressure and temperature~\citep[see Fig. 2 of][]{Tinetti2012}. We note that the water cross-section at $1.4\mu m$ is small compared to the cross-section at longer wavelengths, meaning that our estimate of the photosphere is a conservative upper limit of the photospheric pressure. }

We now assume that the temperature at the photosphere is close to the dayside equilibrium temperature calculated by assuming a zero albedo and a poor, dayside only, atmospheric redistribution of energy. Both the small albedo of ultra hot Jupiters and weak redistribution are theoretical expectations~\citep{Perez-Becker2013a,Komacek2016} that have been confirmed by the observations so far~\citep{Komacek2017,Bell2017}. We then calculate the photospheric abundance of water given the calculated pressure and temperature and compare this value to the expected abundance of water in the hypothetical case of a lack of thermal dissociation. If the fraction of water molecules that are dissociated at the expected photosphere is large, then no self-consistent solution with a constant water abundance can be found, and molecular dissociation likely plays an important role in shaping the spectrum of the planet.

\begin{figure}[h!]
\includegraphics[width=\linewidth]{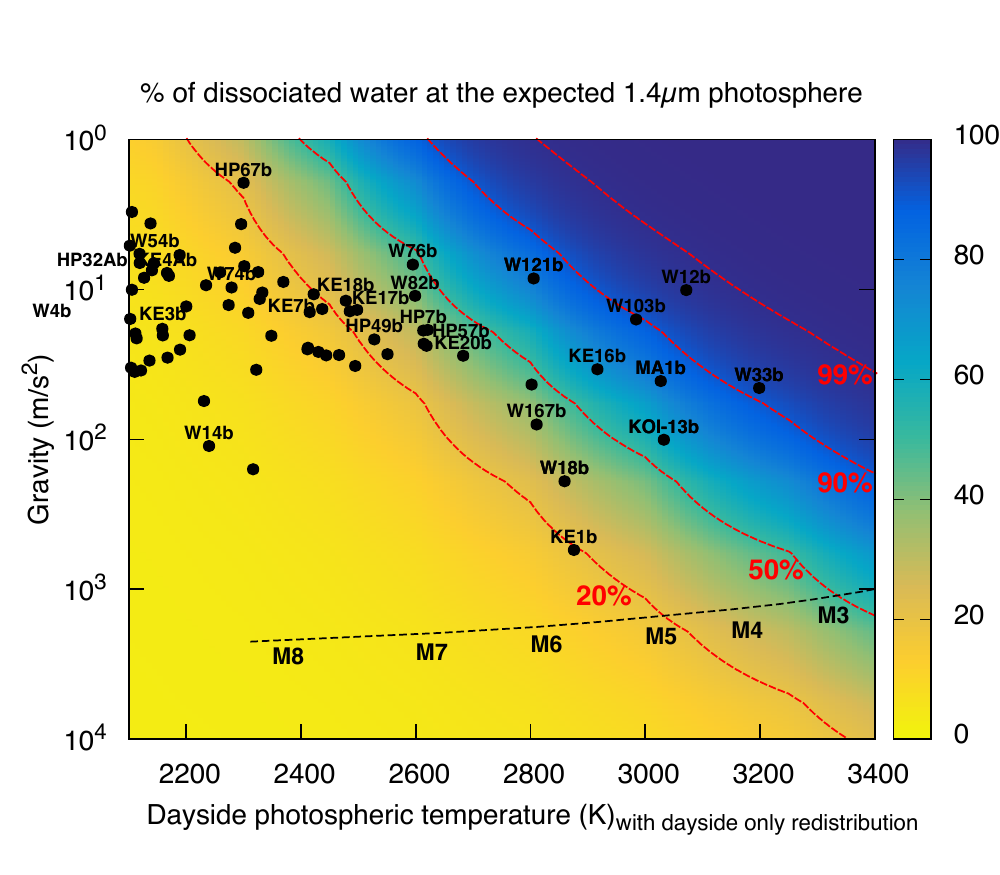}
\caption{Dayside temperature assuming a dayside-only redistribution of the incoming energy \emph{vs.} gravity of the planet for all known transiting planets. The background shows the percentage of water that is dissociated at the $1.4\mu m$ photospheric level calculated assuming water does not dissociate. If a planet is in a significantly dissociated zone of the plot, then water dissociation needs to be taken into account to calculate the emission spectra of the planet. A stellar isochrone for an age of 4\,Gyr from~\citet{Baraffe1998} is shown as a dashed black line. Water dissociation should be important in all planets hotter than the $20\%$ contour line. For clarity, only the name of planets orbiting stars with magnitude V smaller than 10.5 are shown for dayside temperatures lower than 2700K.}
\label{fig:safe}
\end{figure}

In Fig.~\ref{fig:safe}, we show the fraction of water that should be thermally dissociated at the expected photosphere of the planet (i.e., the photosphere calculated by assuming a constant water abundance) as a function of planet dayside temperature (assuming a dayside-only redistribution of energy) and planet gravity. Water dissociation increases with increasing planet temperature and with decreasing planet gravity (planets with a lower gravity have a smaller photospheric pressure). As a result, the fraction of water dissociated at the expected photosphere of these planets follows a tilted line in the gravity-temperature diagram. 

The figure also shows the current planet population together with a stellar isochrone at 4\,Gyr calculated using the evolution models of~\citet{Baraffe1998}. For stars, the water feature at $1.4\mu m$ is significantly damped for M5 and hotter stars~\citep{Kirkpatrick1993}. By comparing with the location of the stellar isochrone in Figure~\ref{fig:safe}, we conclude that water dissociation should play a significant role in shaping the spectral properties of all planets that lay on the right-hand side of the $20\%$ contour line. On top of the four planets studied in detail in this paper, it includes well-studied planets (e.g., WASP-12b, WASP-33b, Kepler-13b) and promising targets for atmospheric characterization (e.g., MASCARA-1b). 

\subsection{Influence on molecular inferences from spectra}

When atmospheric retrievals are performed on a planetary spectrum, the model adjusts the molecular abundances and pressure-temperature profile to find the best fit to the spectrum. To first order, the molecular opacities determine which layers of the atmosphere are probed as a function of wavelengths and the temperature gradient is adjusted to obtain the temperature ratio between the probed atmospheric layers that corresponds to the observed flux. If the molecular dissociation is neglected, as was done in a number of previous retrieval studies, then the same temperature difference would be retrieved over a wider range of pressures. As a consequence, a retrieval framework that does not take into account the presence of a vertical gradient of spectroscopically active molecules would retrieve a smaller temperature gradient than the actual one. 

Neglecting the presence of molecular dissociation and $\rm H^-$ in atmospheric retrievals of ultra hot Jupiters can also bias the retrieved elemental abundances. When constant abundances are assumed, two solutions can explain the lack of features in the $1.1-1.7\mu m$ wavelength range: either the atmosphere is isothermal, and therefore emits like a blackbody at all wavelengths, or the atmosphere has no water in its atmosphere. For a few of these planets, the presence of atmospheric features at $4.5\mu m$ detected by Spitzer rules out the isothermal solution, leading to the conclusion of a carbon-rich, oxygen-poor atmosphere. The presence of molecular dissociation and $\rm H^-$ will likely provide an alternative explanation compatible with an oxygen-rich atmospheric composition: water is not seen in the $1.1-1.7\mu m$ range because of both thermal dissociation and the presence of $\rm H^-$, but CO, which does not dissociate, is seen at $4.5\mu m$.  

Molecular dissociation provides two additional ways to probe the atmospheres of ultra hot Jupiters. First, the difference between the dissociation rate of water and CO allows one to probe very different layers of the atmosphere with a broadband spectrum of the atmosphere, allowing for a better constraint on the pressure-temperature profile from the observations than for a cooler planet where all the contribution functions in the infrared peak at similar pressure levels. 

Second, the difference in the pressure dependence of the different molecules creates a predictable variation of the abundance ratio with pressure. The measured relative molecular abundances (e.g., TiO and water or CO and water) could constrain the pressure level of the photosphere and remove the usual degeneracy between atmospheric metallicity and energy redistribution. Such a constraint, however, would depend on the prior set on the elemental abundance ratio, that is, the relative abundance of TiO and water at the photosphere could be fitted by either changing the relative elemental abundance of titanium and oxygen or by changing the photospheric pressure (and the same applies for the C/O ratio and the relative abundance of water and CO). The pressure dependence of the H$^-$ opacities could probably be used in a more unique way to constrain the photospheric pressure, although it depends in a non-trivial way on the electron abundance, which is a function of the sodium and potassium elemental abundances.

\subsection{Comparison with the observations}
\label{sec:Obs}

\begin{figure*}
\includegraphics[width=\linewidth]{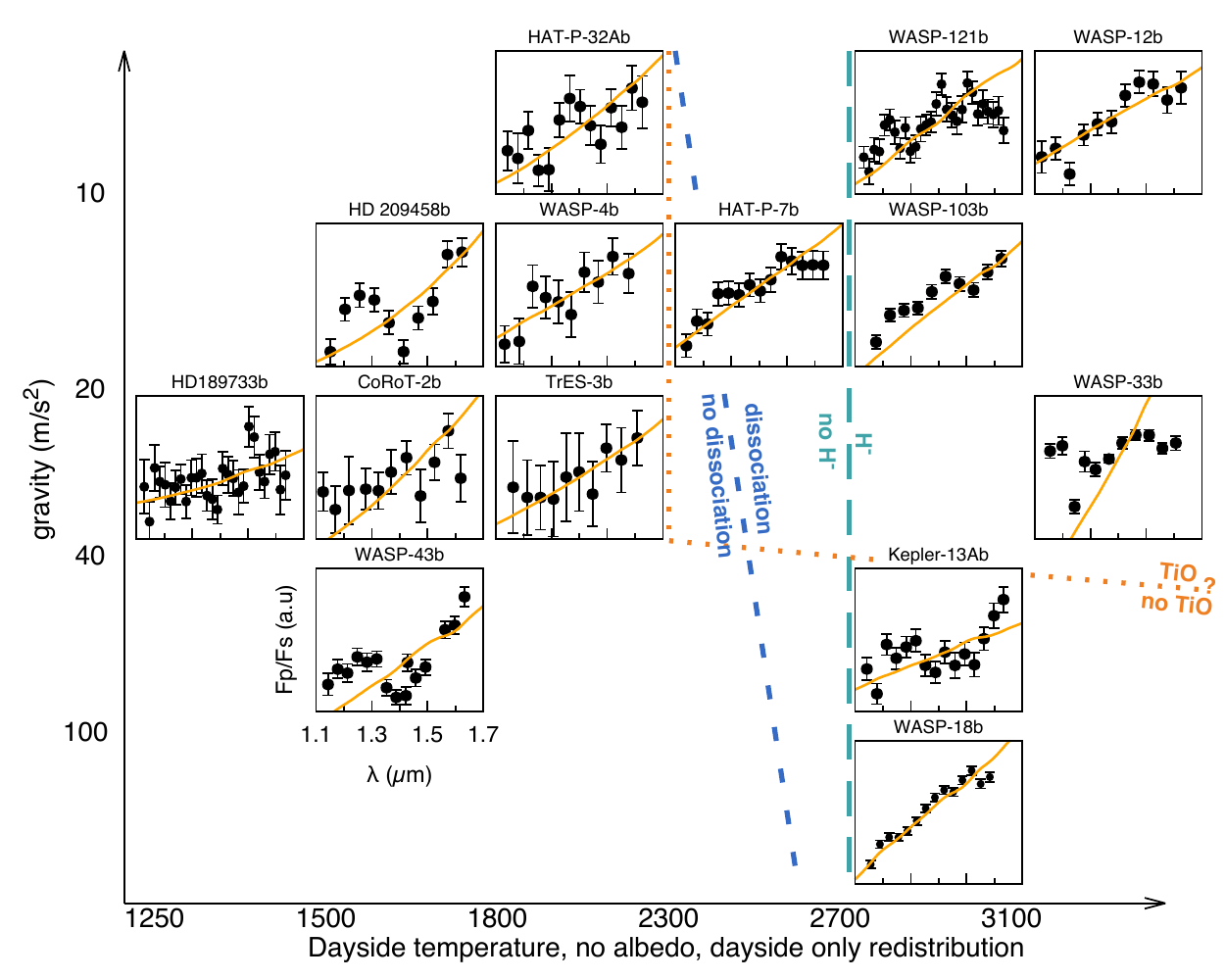}
\caption{Published secondary eclipse spectra of exoplanets observed by HST/WFC3 with the G141 grism. The vertical scale is different for each planet. The spectra have been ordered as a function of gravity and dayside equilibrium temperature assuming zero albedo and a dayside only redistribution of the absorbed stellar energy. $\rm H_2O$ dissociation should be important for planets right of the short dashed line and $\rm H^-$ opacities should be relevant for planets right of the long dashed line. { We additionally show as a dotted line the putative separation between planets with gaseous TiO and planets where TiO is cold trapped either in the deep layers~\citep{Parmentier2016} or in the nightside of the planet~\citep{Parmentier2013,Beatty2017}. In each panel the orange line shows the best fit blackbody, whose temperature is given in Table~\ref{Table:BB}.}}
\label{fig:All-WFC3}
\end{figure*}

\begin{table}
\centering
\caption{Blackbody temperatures based on HST/WFC3 data.}
\label{Table:BB}
\begin{tabular}{c | c | c}
\hline \hline
Planet & WFC3 Data Source & Blackbody Temp. [K]$^a$ \\
\hline
CoRot-2b & \citet{Wilkins2014} & $1705 \pm 15$ \\
HAT-P-7b & \citet{Mansfield2018} & $2693 \pm 14$ \\
HAT-P-32Ab & \citet{Nikolov2018} & $1855 \pm 16$ \\
HD 189733b & \citet{Crouzet2014} & $1420 \pm 4$ \\
HD 209458b & \citet{Line2016a} & $1601 \pm 13$ \\
Kepler-13Ab & \citet{Beatty2017a} & $2886 \pm 19$ \\
TrES-3b & \citet{Ranjan2014} & $1801 \pm 39$ \\
WASP-4b & \citet{Ranjan2014} & $1984 \pm 26$ \\
WASP-12b & \citet{Stevenson2014} & $2894 \pm 19$ \\
WASP-18b & \citet{Arcangeli2018} & $2818 \pm 4$ \\
WASP-33b & \citet{Haynes2015} & $2801 \pm 5$ \\
WASP-43b & \citet{Kreidberg2014a} & $1683 \pm 6$ \\
WASP-103b & \citet{Kreidberg2018} & $2932 \pm 8$ \\
WASP-121b & \citet{Evans2017} & $2650 \pm 10$ \\
\hline

\end{tabular}
\flushleft $^a$ The uncertainties do not take into account the uncertainty in the stellar effective temperature, which can be larger than the uncertainty derived from the data. The use of different stellar parameters/models can be a source of discrepancy with previous studies. 

\end{table}

In Fig.~\ref{fig:All-WFC3}, we show all the published or submitted hot jupiter secondary eclipse spectra taken using the WFC3 instrument on board the Hubble Space Telescope. The data are sorted by gravity and dayside equilibrium temperature assuming zero albedo and dayside-only redistribution of the stellar energy, as in Fig.~\ref{fig:safe}. Since we did not re-analyze the raw data directly, some of the differences between the spectra might be due to the use of different data analysis pipelines, although independent analyses typically yield similarly shaped spectra even if different values are found for the overall secondary eclipse depth \citep[e.g.,][]{Cartier2017,Kreidberg2018}. The dotted line in Fig.~\ref{fig:All-WFC3} shows the approximative location of the $20\%$ H$_{2}$O dissociation boundary of Fig.~\ref{fig:safe}, and the dashed line shows the approximate location where $\rm H^-$ opacities become important at the $1.4\mu m$ photosphere based on Fig.~\ref{fig:PTprofile}. On the right side of the dotted line, water dissociation is expected to play a significant role in shaping the spectrum in the WFC3 bandpass, while on the right-hand side of the dashed line, $\rm H^-$ is expected to play an important role. { In order to complete the picture, we added as a dotted line the putative separation between planets where gaseous TiO is expected to be present in oxygen-rich atmospheres and planets where TiO is expected to rain out of the atmosphere, either due to the deep cold trap~\citep{Spiegel2009,Parmentier2016} or the day/night cold trap~\citep{Parmentier2013,Beatty2017}.}
For planets with dayside temperatures cooler than $2300\,\rm K$, an absorption feature of water can be seen in two planets (HD209458b and WASP-43b), whereas the observations are too noisy to detect spectral features in the other planets. For planets with a dayside temperature higher than $2300\,\rm K$, no spectral features as obvious as the one seen in the cooler planets are present and most of the spectra are close to their best fit blackbody. In three cases, a statistically significant departure from the best fit blackbody is observed.

For WASP-121b, the departure from a blackbody is partly based on the six redward points that fall below the best fit blackbody. Interestingly, for all the hot planets shown in the figure apart for Kepler-13Ab, the reddest points are below the best fit blackbody, indicating a possible common origin, either in the planet atmosphere or in the instrument systematics.
For Kepler-13Ab, the presence of an emission feature was observed by~\citet{Beatty2017a} and taken as evidence for the lack of TiO/VO in the planet's atmosphere, leading the authors to propose that the presence of thermal inversions could correlate with the planetary gravity. Whereas this is in line with the expected role played by gravity in the cold trapping of TiO and VO~\citep{Parmentier2013}, the lack of spectral features in the even higher gravity WASP-18b \citep{Arcangeli2018}, where the day/night cold trap should play a larger role, makes the presence of an absorption feature in the spectrum of Kepler-13Ab unusual. Moreover, extrapolating from our study of WASP-121b (see Fig.~\ref{fig:SpectrumComp}), we predict that solar composition atmospheres without TiO/VO should not have any significant spectral feature in the WFC3 bandpass because of the combination of a very shallow temperature profile and the presence of a large molecular gradient. The absolute temperature gradient could be larger if the stellar light were absorbed at a deeper level. This could be obtained through an elemental depletion of sodium compared to oxygen as this would reduce the two main absorbers of stellar light: sodium directly, and $H^-$ through a reduction in the number of free electrons, but would keep the abundance of water constant, the main molecule responsible for the radiative cooling of the atmosphere. This speculation could be tested directly by observing the strength of the sodium absorption during the transit of Kepler-13Ab. 

The spectrum of WASP-33b is perhaps the most unusual of all, with a strong departure from a blackbody spectrum at most wavelengths that was interpreted as emission bands from TiO by~\citet{Haynes2015}. Based on Fig.~\ref{fig:safe}, molecular dissociation should be as important in WASP-33b as it is for WASP-103b. The TiO and water features should be significantly damped due to molecular dissociation and the presence of H$^-$ in a solar-composition model.

WASP-12b deserves a special mention. Although the planet has a spectrum very close to a blackbody in the WFC3 bandpass, it also shows a strong absorption feature at $4.5\mu m$. This led \citet{Madhusudhan2011b} and \citet{Stevenson2014} to conclude that the atmosphere was indeed water poor due to a high elemental ratio of carbon versus oxygen. This interpretation was challenged by \citet{Kreidberg2015}, who measured a solar water abundance at the limb of the planet. While thermal dissociation provides a natural explanation for the blackbody-like shape in the WFC3 bandpass and the different retrieved abundances between the dayside and the limb of the planet, our solar-composition model does not predict absorption features in the dayside at $4.5\mu m$, even with a TiO depleted atmosphere.

\section*{Conclusions}

Ultra hot Jupiters with dayside temperatures larger than $2200\,\rm K$ are good targets for thermal emission measurements. However, the majority of the observed planets have weaker-than-expected spectral features in the $1-2\mu m$ range. Using the example of WASP-121b, we interpret this lack of strong features as being due to a combination of a vertical gradient in molecular abundances due to thermal dissociation, and to the presence of $\rm H^-$ absorption at wavelengths shorter than $1.4\mu m$.

Thermal dissociation affects all spectrally important molecules in the atmospheres of ultra hot Jupiters except CO. It creates a large vertical gradient in the molecular abundances. We show analytically that the presence of such a molecular gradient weakens the features in emission spectra. This is a qualitatively different effect than a global depletion of the abundances.

The vertical abundance gradient in the atmospheric regions dominated by thermal dissociation is different for different molecules. As a consequence, the abundance ratios of different molecules vary with height. Specifically, we show that the abundance ratio between TiO and H$_2$O increases with decreasing pressure, meaning that planets with a photosphere at lower pressures (e.g., atmospheres with a higher metallicity and/or lower-gravity planets) should have a sharper thermal inversion. As a consequence, the shape of the pressure temperature profile becomes more sensitive to the metallicity, which could be exploited by retrieval models to more precisely constrain the atmospheric metallicity of ultra hot Jupiters. { In such an approach, however, the metallicity might be degenerate with the elemental abundance ratio Ti/O and caution is required when interpreting the observations.}

Similarly, the abundance ratio between Na and H$_2$O increases with decreasing pressure. As a consequence, even when TiO and VO are removed from the atmosphere, our solar composition models still have dayside thermal inversions that are caused by the absorption of stellar light by Na at pressure levels where the H$_2$O bands are optically thin. The resulting pressure-temperature structure is a combination of inverted and non-inverted profiles, leading to the presence of absorption features of H$_2$O and emission features of CO in the same spectrum.

Carbon monoxide does not dissociate in the atmospheres of currently known ultra hot Jupiters (with the exception of KELT-9b), meaning that the H$_2$O and CO bands probe very different atmospheric levels. The low-resolution spectra of ultra hot Jupiters therefore contain information on the thermal structure of the planets' atmospheres over a wider range of pressures than they do for cooler planets. Observations of both CO and H$_2$O bands using, for example, the G395 grating of JWST/NIRSPEC in the $3-5.5\mu m$ range would allow a precise constraint on the radiative balance of these planets' atmospheres at different pressure levels.

{ Given the large horizontal temperature contrasts predicted for the atmospheres of ultra hot Jupiters, molecular abundances are expected to vary by orders of magnitude between the dayside, the limb, and the nightside of the planet. While the secondary eclipse spectrum probes a water-depleted atmosphere, the transmission spectrum probes a part of the atmosphere where water has recombined.}

For a model of WASP-121b without drag, in which the hot spot is shifted eastward, the water-depleted hot spot dominates the spectrum from phase $-60^{\circ}$ to $120^{\circ}$. We predict that the planet spectrum should only show large absorption features of water from phases ranging from $-120^{\circ}$ to $-180^{\circ}$. However, these nightside features could be damped by the presence of equilibrium condensate clouds. 

We re-interpret the transmission spectrum of WASP-121b as being consistent with a solar composition atmosphere having partial cloud coverage. We predict that $\rm CaTiO_3$ clouds should form in the west limb between $1\,\rm bar$ and $1\,\rm mbar$, damping the water feature observed by HST/WFC3. Gaseous TiO would still be seen above the clouds on the west limb and over the whole east limb, creating a strong absorption feature in the optical. TiO would therefore be thermally dissociated on the planet's dayside, partially condensed on the atmospheric limb, and fully condensed on the planet's nightside. For this scenario to be realistic, the recombination timescale of gaseous TiO and the condensation timescale of Ti-bearing condensates must be short compared to the atmospheric circulation timescale. Moreover, the atmospheric mixing must be strong enough to avoid the depletion of TiO through the day/night cold trap mechanism. 

Finally, we consider the role of H$_2$O dissociation in the currently known population of ultra hot Jupiters. The importance of H$_2$O dissociation in the atmospheres of substellar objects is expected to increase with photospheric temperatures and decrease with increasing gravity. As a consequence, M dwarf stars with similar photospheric temperatures to hot Jupiters are less sensitive to the effect of H$_2$O dissociation. 

Water dissociation should be important for all the planets that are on the right-hand side of the $20\%$ line of Fig.~\ref{fig:safe}. Among them are targets with existing atmospheric characterization such as WASP-12b, WASP-33b, WASP-103b, WASP-121b, Kepler-13Ab, KELT-1b, WASP-18b, and HAT-P-7b. For all these planets, atmospheric retrieval studies that have been performed without taking into account the important role of $\rm H^-$ opacities and the presence of a vertical gradient of the H$_2$O abundance might have obtained biased inferences of the elemental abundances, elemental abundance ratios, and pressure-temperature profiles. Particularly, all the inferences of a high C/O ratio based on the comparison between blackbody-like emission spectra in the WFC3 bandpass and the presence of absorption or emission features in the Spitzer/IRAC bandpasses will likely be challenged when considering the presence of non-constant molecular abundances.

In conclusion, although our theoretical work provides a good explanation for the lack of strong molecular features in many ultra hot Jupiters, we are unable to explain the strong and diverse departures from a blackbody spectrum seen in WASP-121b, WASP-33b and Kepler-13Ab. Observations of a larger sample of ultra hot Jupiters are needed to understand their diversity and correctly constrain their compositions. Future observations with HST/WFC3 should target planets orbiting bright stars to achieve a high-enough signal-to-noise ratio to detect the weaker-than-usual molecular features. In the long term, our understanding of these atmospheres should improve significantly by obtaining emission spectra with JWST at longer wavelengths where molecular features are expected to be larger.

\begin{acknowledgements}
The work in this paper is related to observations obtained by our team with HST (programs GO-13467, GO-14050, and GO-14792) and Spitzer (program 11099). Support for the HST programs was provided by NASA through a grant from the Space Telescope Science Institute, which is operated by the Association of Universities for Research in Astronomy, Inc., under NASA contract NAS 5-26555. Support for the Spitzer program was provided by NASA through an award issued by JPL/Caltech. J.L.B. acknowledges support from the David and Lucile Packard Foundation. R.L. acknowledges support from the NASA XRP program. J.M.D. acknowledges that the research leading to these results has received funding from the European Research Council (ERC) under the European Union's Horizon 2020 research and innovation programme (grant agreement no. 679633; Exo-Atmos). M.R.L acknowledges that the research leading to these results has received funding from the NASA XRP grant NNX17AB56G and also acknowledges the ASU Research Computing staff for support with the Saguaro and Agave compute clusters. 
\end{acknowledgements}

\bibliography{Parmentier2018.bbl}
\end{document}